\documentclass{timproc}

\usepackage[utf8]{inputenc}
\usepackage{amsmath}
\usepackage{amsfonts}
\usepackage{amssymb}
\usepackage{graphicx}

\usepackage[figuresright]{rotating}
\usepackage{multirow}

\usepackage{hyperref}

\title[Hydrodynamic scaling of collisions in different mass regimes]{Applicability and limits of simple hydrodynamic scaling for collisions of water-rich bodies in different mass regimes}

\author[Burger \& Sch{\"a}fer]{C. Burger$^1$ and C. M. Sch{\"a}fer$^2$}
\affiliation{$^1$University of Vienna, Department of Astrophysics, T{\"u}rkenschanzstra{\ss}e 17,\\ 1180 Vienna, Austria\\
	\email{c.burger@univie.ac.at}\\[1ex]
$^2$Institut f{\"u}r Astronomie und Astrophysik, Eberhard Karls Universit{\"a}t T{\"u}bingen, Auf der Morgenstelle 10, 72076 T{\"u}bingen}

\begin{document}
\setcounter{page}{63}

\maketitle

\begin{abstract}
We investigate the outcome of collisions in very different mass regimes, but an otherwise identical parameter setup, comprising the impact velocity ($v/v_\mathrm{esc}$), impact angle, mass ratio, and initial composition, w.r.t.\ simple hydrodynamic scaling.
The colliding bodies' masses range from $\simeq 10^{16}$ to $10^{24}$\,kg, which includes km-sized planetesimals up to planetary-sized objects.
Our analysis of the results comprises the time evolution of fragment masses, the fragments' water contents and fragment dynamics, where we start with bodies consisting of basalt and water ice.
The usual assumption of hydrodynamic scaling over a wider range of masses is based on material behavior similar to a fluid, or a rubble pile, respectively. All our simulations are carried out once including full solid-body physics, and once for strengthless -- but otherwise identical -- bodies, to test for the influence of material strength.

We find that scale-invariance over a wider range of masses is mostly only a very crude approximation at best, but can be applied to constrained mass ranges if tested carefully.
For the chosen scenarios the outcomes of solid-body objects compared to strengthless fluid bodies differ most for our intermediate masses, but are similar for the lowest and highest masses.
The most energetic, planet-sized collisions produce considerably faster and more fragments, which is also reflected in high water losses -- above 50\% in a single collision.

\keywords{hydrodynamics -- methods: numerical -- planets and satellites: formation}
\end{abstract}

\section{Introduction}
Collisions on vastly different size scales are an ubiquitous process during all stages of planet formation, and occasionally still occur in our Solar System today.
The current understanding of planet formation suggests that after planetesimals and planetary embryos formed on relatively distinct orbital distances from the central star, a final phase -- late-stage accretion -- starts, which is characterized by chaotic interactions and thus also violent collisions. It is believed that the final characteristics of planets are predominantly set during their late formation history, by a relatively small number of decisive events. While the composition of planetesimals and embryos originally reflects their birth region, some material is now scattered throughout the system. This is of particular interest for radial mixing of volatile material, especially water, which can be transported into the (inner) terrestrial planet region \cite[see e.g.][for a comprehensive overview]{Raymond2014_Terrestrial_formation_at_home_and_abroad}.
These processes are typically investigated by gravitational N-body simulations, where usually perfect inelastic merging is assumed whenever two bodies collide. This disregard of the wide variety of collision outcomes falsifies results to some degree in general, and particularly for volatiles.
A number of studies aiming for a more accurate treatment of collisional fragmentation came up in recent years \citep{Kokubo2010_Formation_under_realistic_accretion,Genda2012_Merging_criteria_for_giant_impacts,Leinhardt2012_Collisions_between_gravity-dominated_bodies_I,Stewart2012_Collisions_between_gravity-dominated_bodies_II,Marcus2010_Icy_Super_Earths_max_water_content,Chambers2013_Late-stage_accretion_hit-and-run_fragmentation,Quintana2016_Frequency_giant_impacts_Earth-like_worlds}, but especially their treatment of volatile material is still very rudimentary, if included at all \citep[but see e.g.][this volume]{Marcus2010_Icy_Super_Earths_max_water_content, DvorakMaindlBurger2015_Nonlinear_phenomena_complex_systems_article, Maindl2017_This_volume}.

Collisions are complex phenomena. Most of those that occur during planet formation are not directly accessible via laboratory experiments -- which are limited to cm-sized bodies at most -- but can be investigated in detail only by means of numerical simulations.
In the limit of inviscid, incompressible, self-gravitating and strengthless bodies, collisions would be scale-invariant \citep{Asphaug2010_Similar_sized_collisions_diversity_of_planets}, meaning that outcomes would be indistinguishable when lengths were scaled to the bodies' radii, and velocities to $v_\mathrm{esc}$ (see footnote below).
For equal input parameters (in terms of these scaled quantities) this would result e.g.\ in identical orbits, meaning for example that the initial approach time until collision would be the same, independent of the bodies' sizes.
However, real bodies are neither inviscid nor incompressible. Whether they can be treated as strengthless fluid depends on their internal structure, their thermal and collisional history and their size.
Larger bodies can retain heat for longer timescales, and therefore likely possess at least partially molten interiors. Small bodies on the other hand are probably often rather rubble piles than solid rocks or ices, since their masses are too small to compress material in their interiors, and collisions do not deliver sufficient energy for melting -- but continuously shatter what is left of solid material.
In addition material strength becomes generally less important with size when one transitions from the strength-dominated to the gravity-dominated regime.
Yet another important physical aspect are shocks, which arise when passing from subsonic to supersonic collisions, and the related energy/heat deposition.
Nevertheless studies have shown that several aspects of collisions are indeed approximately scale-invariant. An example is the transition between merging and hit-and-run events, where \citet{Genda2012_Merging_criteria_for_giant_impacts} found that for protoplanets this transition does not depend on the total mass (for other parameters equal).

We investigate a typical collision geometry for a wide range of masses, but keep all other parameters constant. The Smooth Particle Hydrodynamics (SPH) simulations are performed twice for each scenario (i.e.\ each total mass), once including full solid-body physics and also for strengthless fluid objects. The properties of real bodies can be expected to lie somewhere between these two limiting cases, thus they can be regarded as confining the results of real collisions in some sense.
Our aim is to exemplarily show differences in collision outcomes on different size scales by focusing on one particular collision geometry. We investigate to what extent and up to what accuracy outcomes behave similar for different-size collisions, including volatile material (water). This is also important for assessing whether or not the total mass needs to be incorporated into the parameter space of future studies.

The two physical models we used are outlined in Sect.~\ref{sect:physical_models}, and the simulation scenarios are described in Sect.~\ref{sect:simulation_scenarios}. We present results in Sect.~\ref{sect:results} and discuss them and conclude in Sect.~\ref{sect:discussion_and_conclusions}.

\section{Physical models}
\label{sect:physical_models}
In this section we outline the most important aspects of the physical models used for the SPH simulations. For a full and comprehensive overview we refer the reader to \citet{Schaefer2016_miluphcuda}.
In the following the continuum mechanics equations describing solid-body physics are summarized, where greek indices denote spatial coordinates and the Einstein summation convention is used.

Conservation of mass is expressed by the continuity equation in Lagrangian form,
\begin{equation}
\label{eq:conservation_of_mass}
\frac{\mathrm{d} \varrho}{\mathrm{d}t} + \varrho \frac{\partial v^\alpha}{\partial x^\alpha} = 0\ ,
\end{equation}
with density $\varrho$ and velocity $v^\alpha$.
Conservation of momentum is given by
\begin{equation}
\label{eq:conservation_of_momentum}
\frac{\mathrm{d} {v^\alpha}}{\mathrm{d}t} =  \frac{1}{\varrho}  \frac{\partial \sigma^{\alpha
\beta} }{\partial x^\beta}\ ,\quad \sigma^{\alpha \beta} = -p \delta^{\alpha \beta} + S^{\alpha \beta}
\end{equation}
with the stress tensor $\sigma^{\alpha\beta}$, pressure $p$ and deviatoric stress tensor ${S}^{\alpha \beta}$ ($\delta^{\alpha \beta}$ is the Kronecker delta).
Conservation of specific energy is finally expressed via
\begin{equation}
\label{eq:conservation_internal_energy}
\frac{\mathrm{d} e }{\mathrm{d} t} = - \frac{p}{\varrho} \frac{\partial v ^\alpha}{\partial x^\alpha} + \frac{1}{\varrho}
S^{\alpha \beta} \dot{\varepsilon}^{\alpha \beta}\ ,
\end{equation}
where the strain rate tensor reads
\begin{equation}
\label{eq:strain_rate_tensor}
\dot{\varepsilon}^{\alpha \beta} = \frac{1}{2} \left( \frac{\partial v^\alpha}{\partial x^\beta} + \frac{\partial
v^\beta}{\partial x^\alpha} \right)\ .
\end{equation}
For elastic deformations the time evolution of the deviatoric stress tensor is specified by Hooke's law.
Plastic behavior is modeled by limiting the deviatoric stress following the von Mises yielding criterion \citep{vonMises1913,Benz1994_Impact_sims_with_fracture_I}.

To model material fracture beyond plastic deformation we use a model for brittle failure introduced by \citet{Benz1995_Sims_brittle_solids_SPH} to SPH.
Once local strains grow large enough a material gets \emph{damaged}, with the scalar value $D \in [0,1]$ indicating the amount of fracture. $D=0$ represents a largely intact region, while $D=1$ indicates fully damaged material that feels no tensile or shear stresses at all.
In a damaged region the elastic stress is reduced via
	\begin{eqnarray}
	\begin{aligned}
	\sigma^{\alpha \beta}_d =& -p_d\, \delta^{\alpha \beta} + (1-D)\, S^{\alpha \beta} \\
	& p_d = \left\{	\begin{array}{c} p	\\ (1-D)\, p	\end{array} \right.
	\quad
	\begin{array}{l}	p\ge 0	\\	p<0	\end{array}
	\ ,
	\end{aligned}
	\end{eqnarray}
thus all components except for positive pressure are reduced by ($1-D$).
Since for material with $D=1$ only positive scalar pressures remain it behaves like a strengthless fluid.

To model thermodynamic behavior the widely used Tillotson equation of state is used \citep{Tillotson1962_Metallic_EOS,Melosh1989_Impact_cratering}, with material parameters from \citet{Benz1999_Catastrophic_disruptions_revisited}.
All simulations include self-gravity, utilizing a tree-code with a $\vartheta$-criterion of 0.5 \citep[see][]{Schaefer2016_miluphcuda}.

We applied this physical model in all our simulations, but for modeling solid bodies we start simulations with $D=0$ (initially), while for strengthless bodies they are already initialized with $D=1$ for all SPH particles.
We will refer to these two models as \emph{solid} and \emph{hydro}, respectively.

\section{Simulation scenarios}
\label{sect:simulation_scenarios}
	\begin{figure}
	\centering
	\includegraphics[width=0.55\hsize]{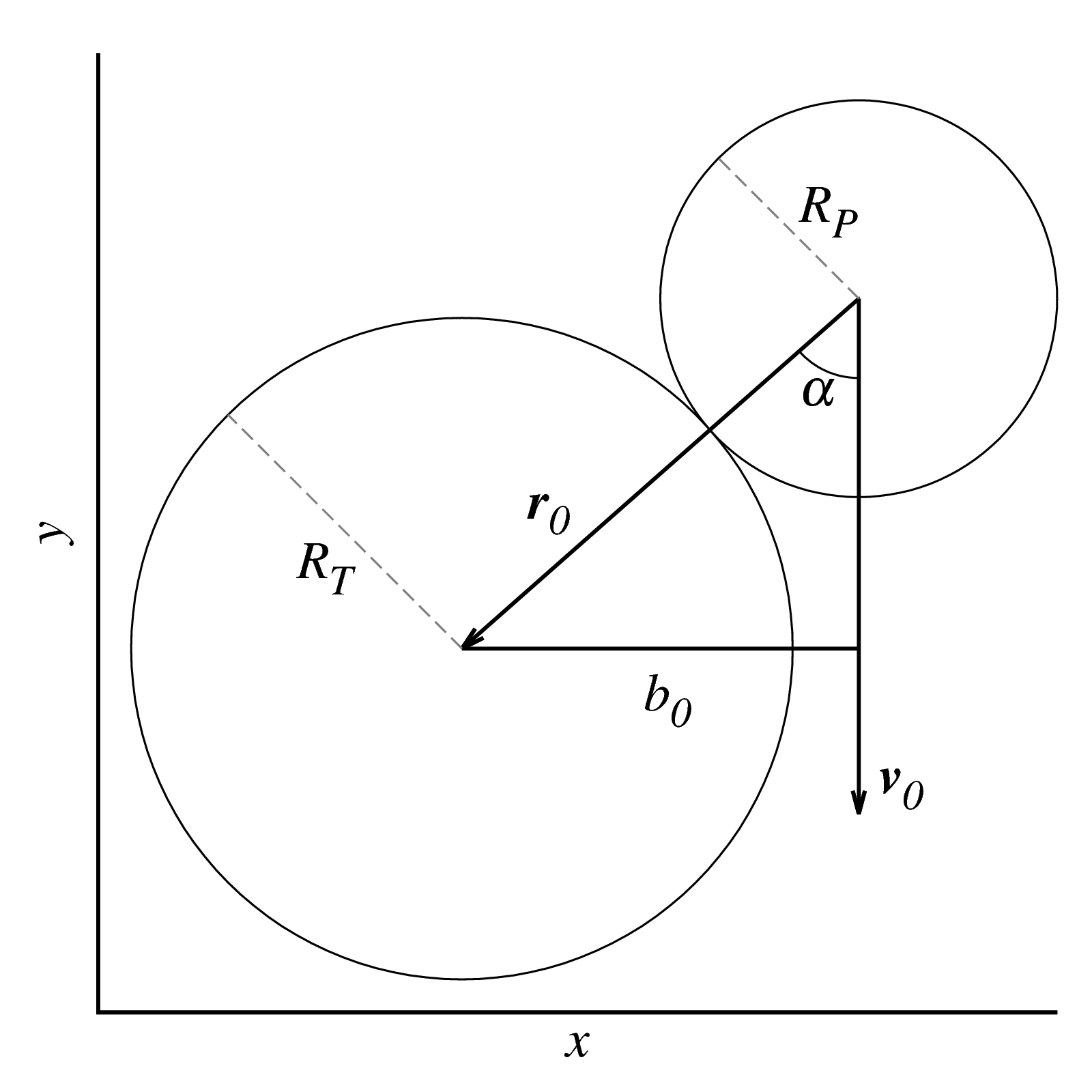}
	\caption{Collision geometry in the rest frame of the target (larger body). Labeled are the impact angle $\alpha$, impact parameter $b_0$, impact velocity vector $\mathbf{v}_0$ and the relative vector from the projectile's center to the target's center $\mathbf{r}_0$. The actual impact velocity is $v_0 = |\mathbf{v}_0|$.}
	\label{fig:collision_geometry}
	\end{figure}
	\begin{figure}
	\centering
	\includegraphics[width=0.49\hsize]{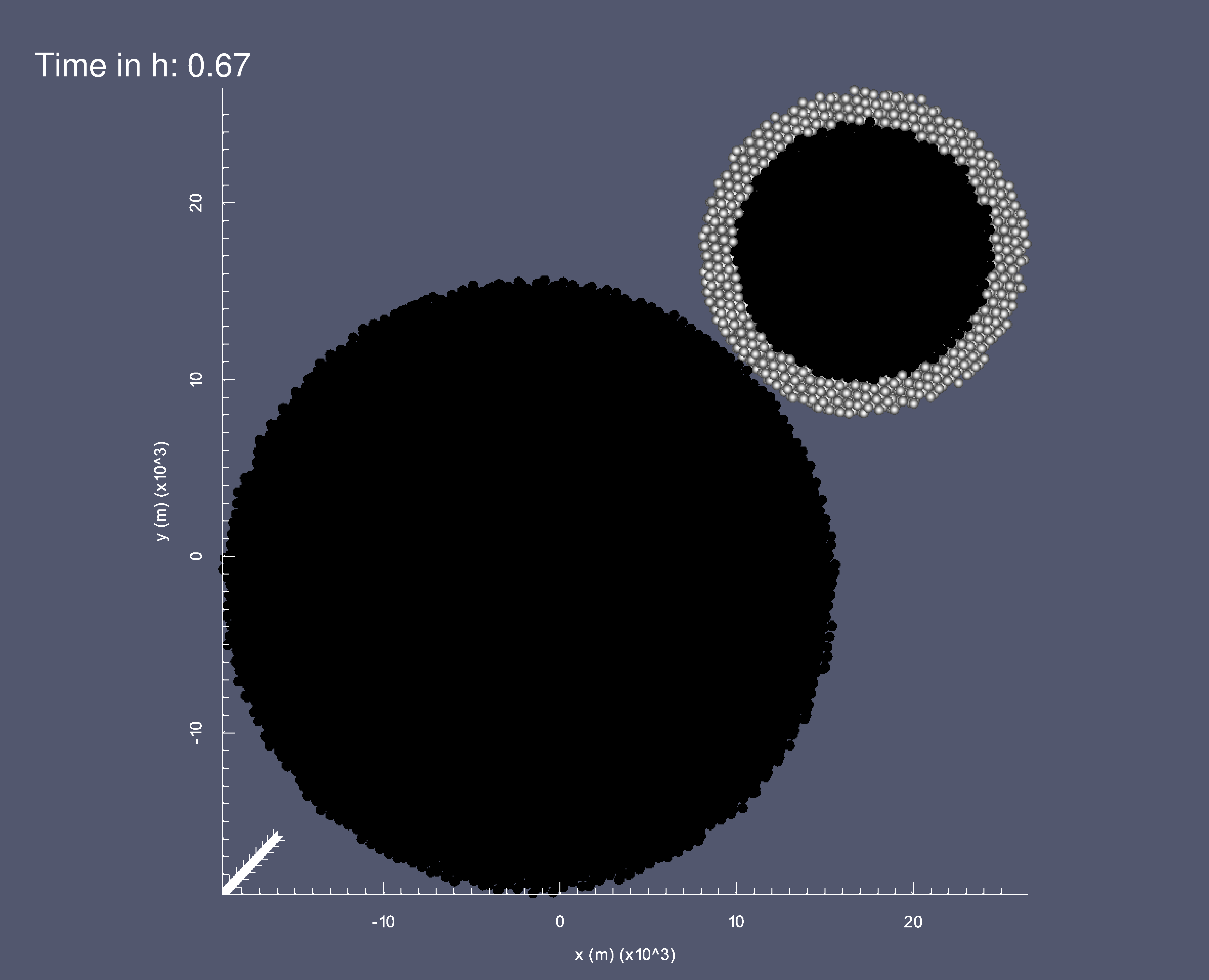}
	\includegraphics[width=0.49\hsize]{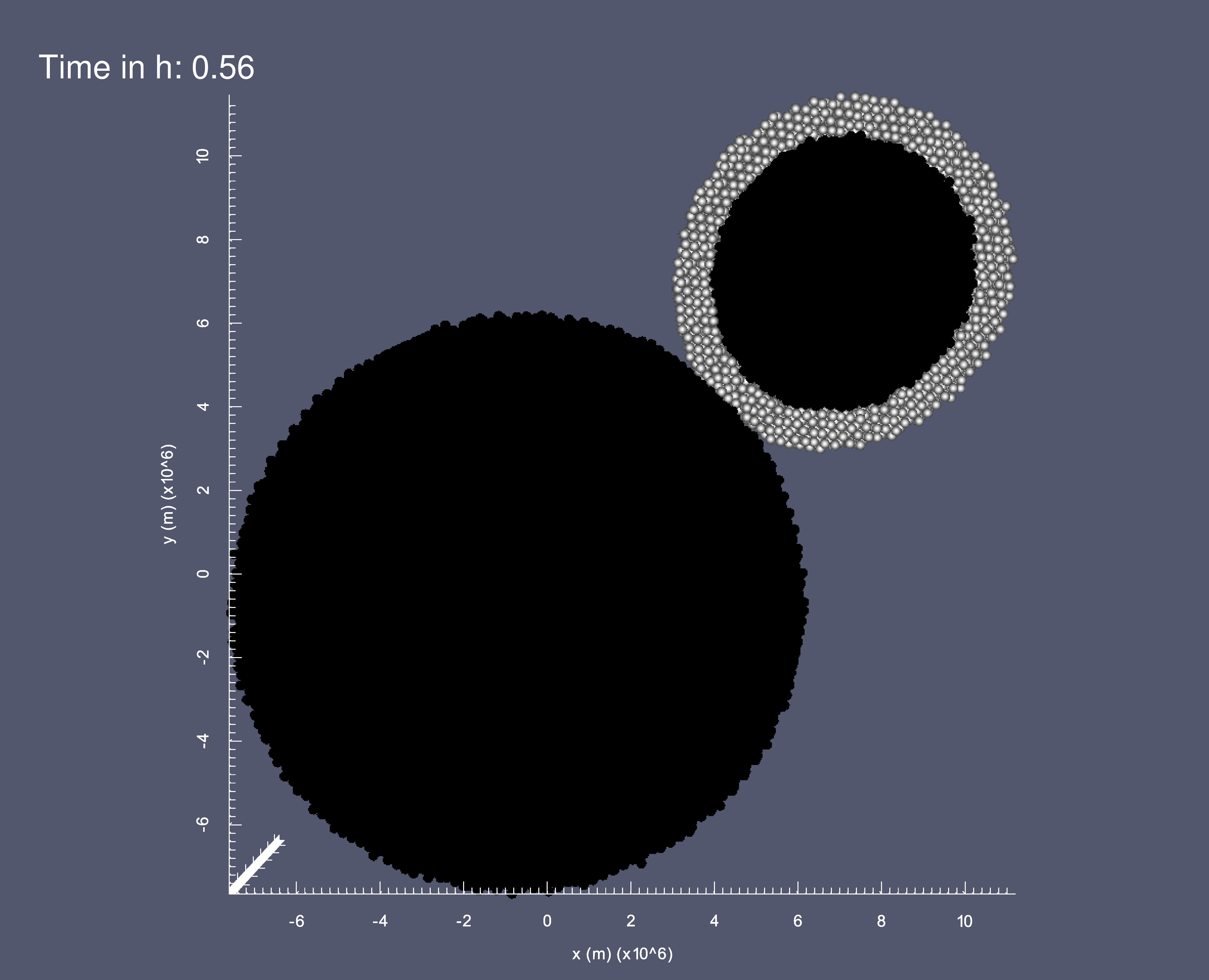} \\
	\vspace{0.2mm}
	\includegraphics[width=0.49\hsize]{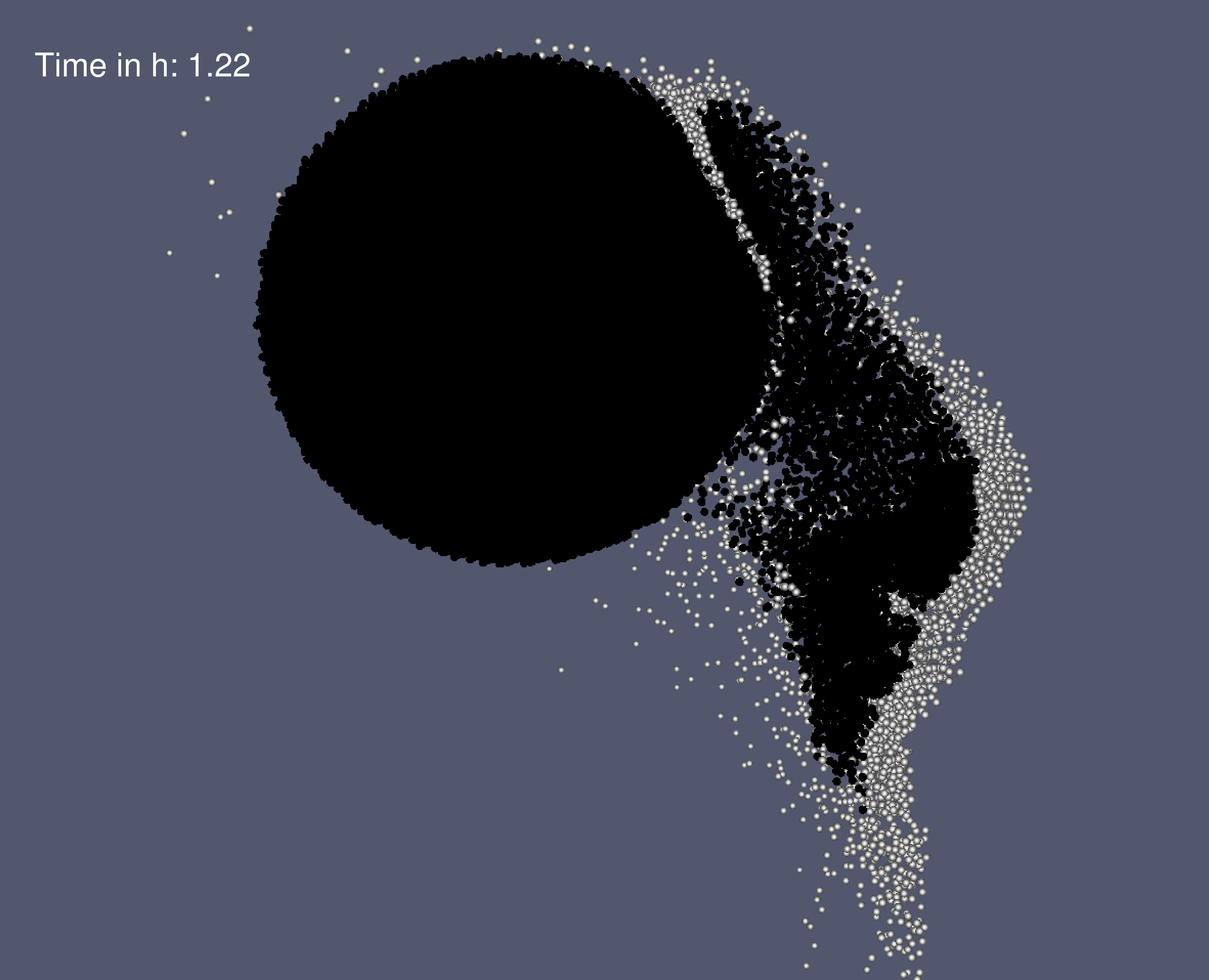}
	\includegraphics[width=0.49\hsize]{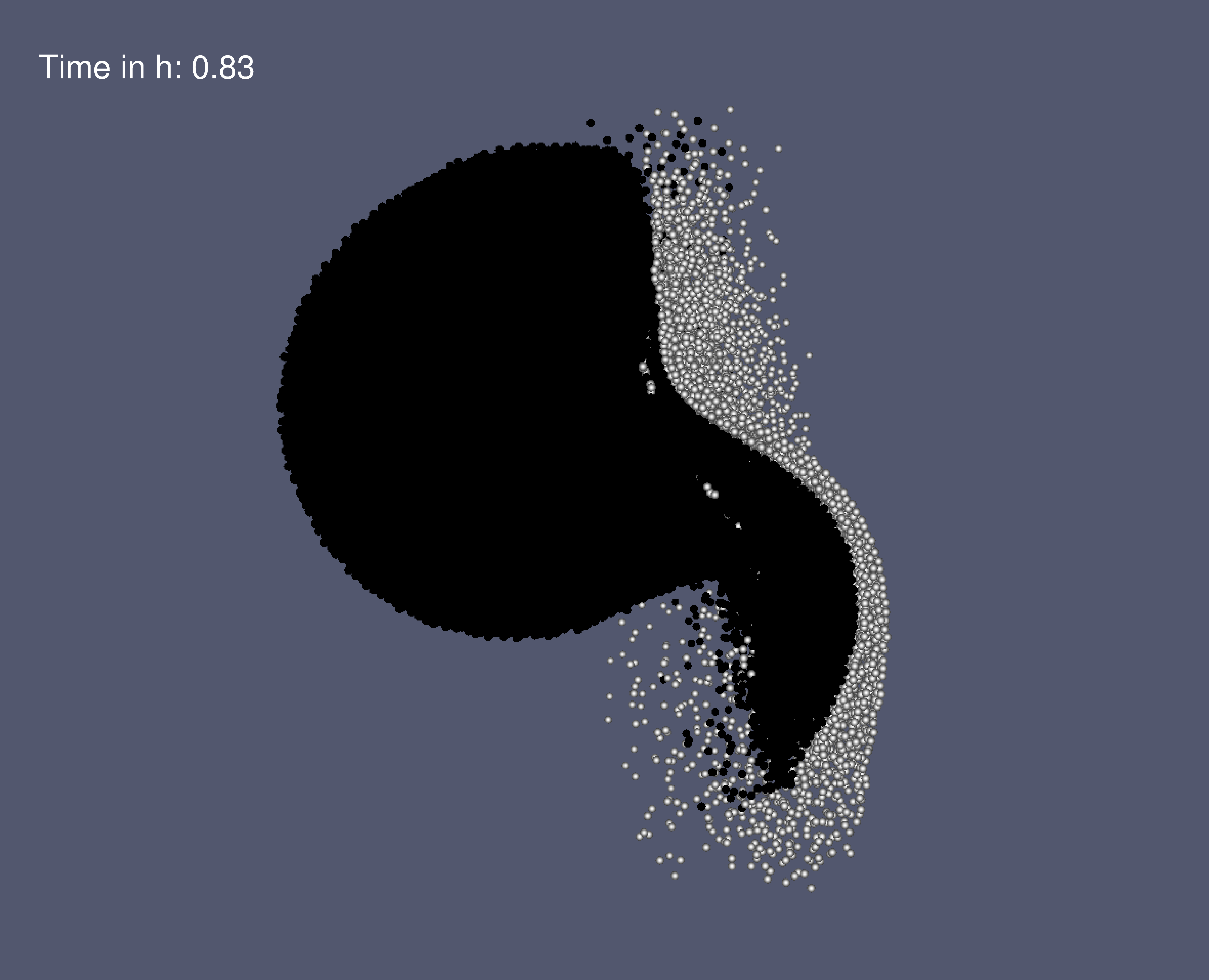} \\
	\vspace{0.2mm}
	\includegraphics[width=0.49\hsize]{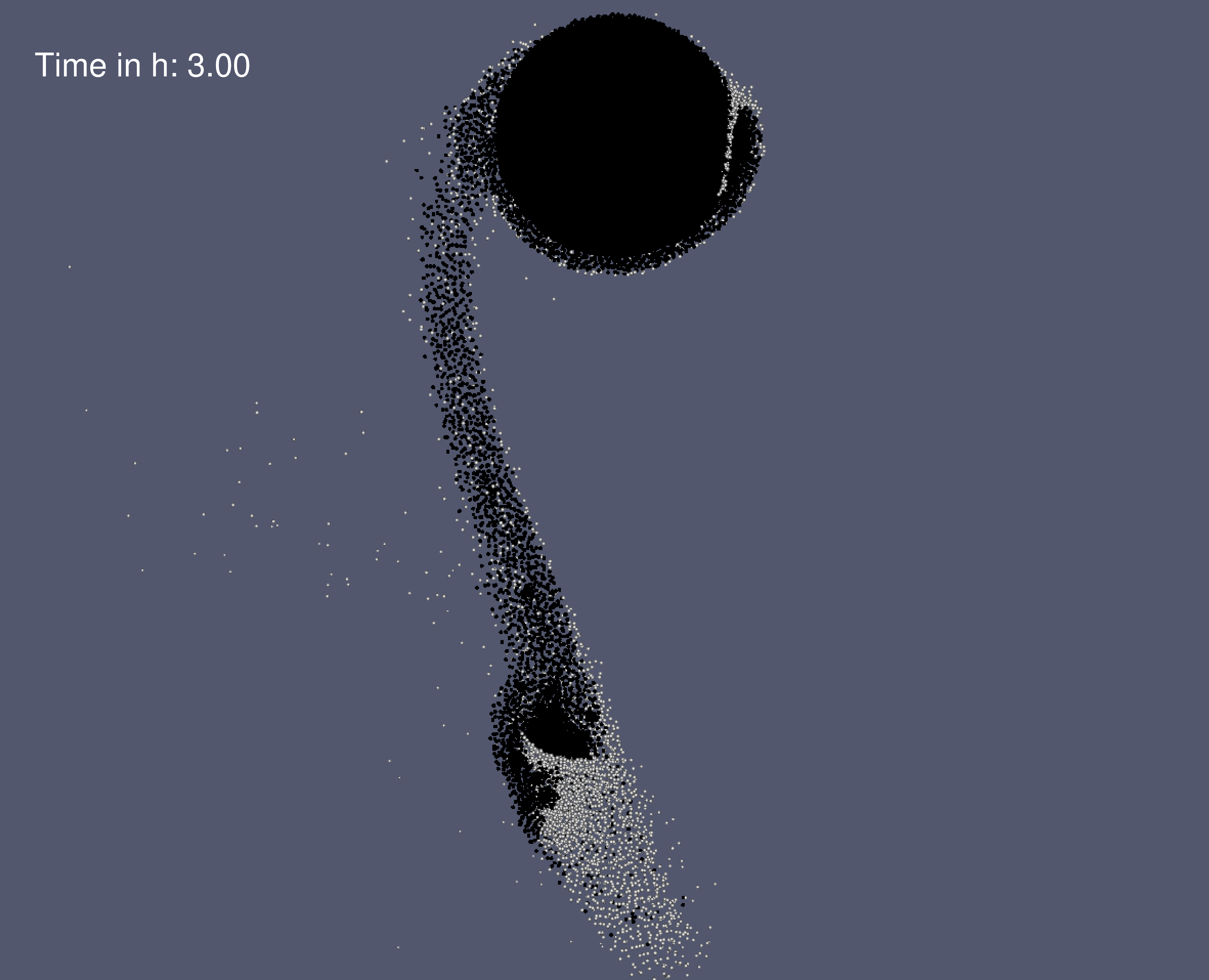}
	\includegraphics[width=0.49\hsize]{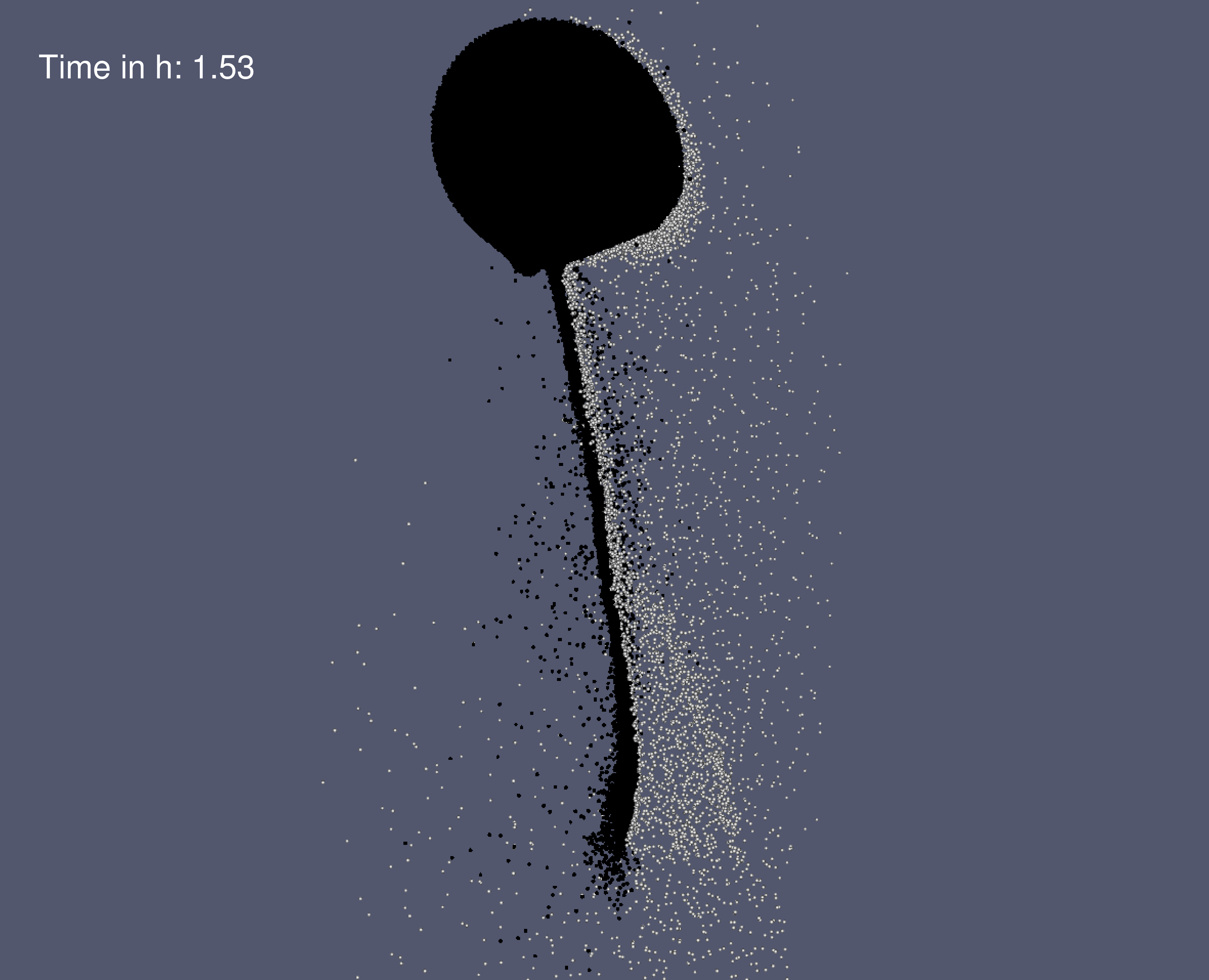}
	\caption{Comparison of simulation snapshots (bodies cut into halves) of the least-massive ($M_\mathrm{tot} = 6.6\times 10^{16}\, \mathrm{kg}$, left panels) and the most-massive ($M_\mathrm{tot} = 6.6\times 10^{24}\, \mathrm{kg}$, right panels) scenario. Material types are color-coded in black (basalt) and white (water ice). These images show the scenarios' respective solid versions, but their hydro counterparts look very similar for these particular scenarios. The times were chosen such that the respective snapshot pairs show roughly similar states during the collision event (see the text for further discussion).}
	\label{fig:16vs24}
	\end{figure}
The chosen scenarios are identical in terms of the projectile-to-target mass-ratio of 1:10, the impact angle $\alpha = 40^\circ$ and a collision velocity of $1.5$ times the mutual escape velocity\footnote{The two-body escape velocity (masses $m_1$ and $m_2$) is given by $v_\mathrm{esc} = \sqrt{2\,G\,M/r}$ with the gravitational constant $G$, $M = m_1+m_2$ and the masses' separation $r$.}.
The impact angle and velocity are both defined at the moment of first contact, as illustrated in Fig.~\ref{fig:collision_geometry}, assuming spherical objects.
We use differentiated bodies where the projectile always consists of a basalt core and a water ice shell of 25 mass-\%, while the target is pure basalt.
These parameters result in a rather grazing collision, where parts of the projectile \emph{miss} the target's cross-section and reaccumulate into a second large fragment after the main collision (see Fig.~\ref{fig:16vs24}).
In all simulations the mass of the target is $6\times 10^\beta$\,kg, with $\beta = $ 16, 18, 20, 21, 22, 23 and 24, thus the most-massive target has approximately one Earth-mass (the projectile always has one-tenth the target-mass).
We will from now on refer to specific scenarios only by the value of $\beta$ for brevity.
Each of these scenarios was simulated once with the full solid-body physics model -- representing e.g.\ molten, differentiated and solidified planetesimals or embryos -- and also with the \emph{hydro} physics model (see Sect.~\ref{sect:physical_models}), yielding 14 simulation runs altogether. We used a total of 125\,000 SPH particles per scenario, which we consider sufficient for resolving the collisions for the intended analysis.
The initial bodies were relaxed by setting their internal structure following self-consistent hydrostatic profiles, with internal energies from (purely) adiabatic compression. They were initially placed apart at 3 times the sum of their radii to allow for the possible build-up of tidal effects.

\section{Results}
\label{sect:results}
Our analysis of simulation results focuses on fragmentation behavior in general, and in particular on the fate of water reservoirs.
All scenarios behave dynamically similar over the course of the simulations, meaning that after the main collision two main fragments are formed again, initially connected by a material bridge which is continuously reaccreted or dispersed as the two main fragments move apart (cf.\ Fig.~\ref{fig:solid_hydro}).
	\begin{figure}
	\centering
	\includegraphics[width=0.6\hsize]{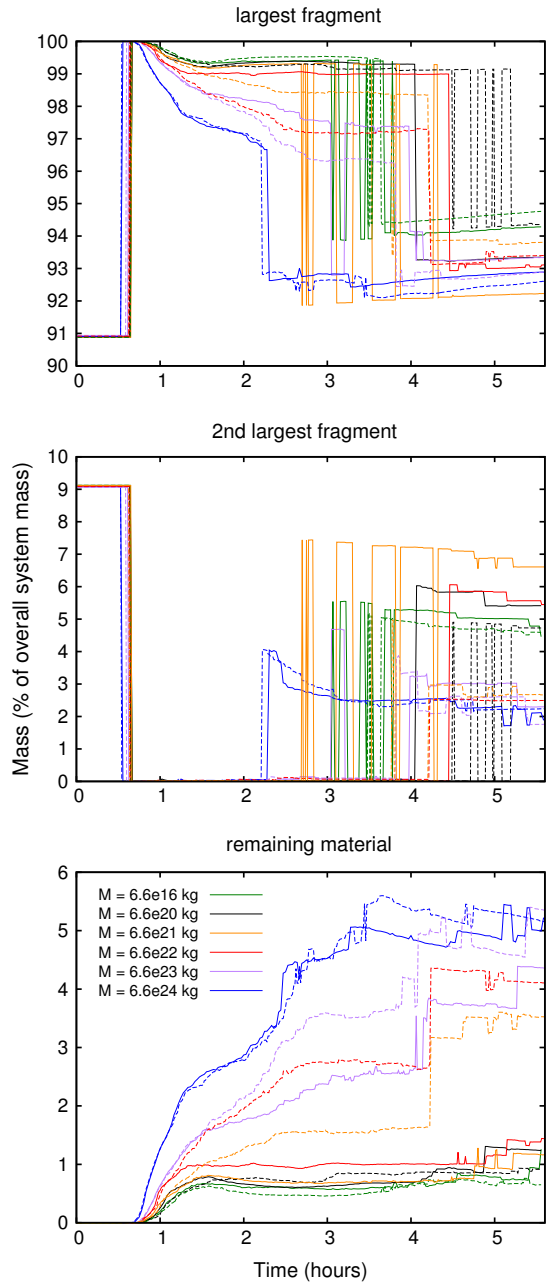}
	\caption{Evolution of the overall mass in the largest fragment, the $2^\mathrm{nd}$ largest one, and the remaining material. Solid runs are indicated by solid lines and hydro runs by dashed lines. Figure~\ref{fig:fragment_masses_ice} shows similar plots for the water ice content.}
	\label{fig:fragment_masses_overall}
	\end{figure}
	\begin{figure}
	\centering
	\includegraphics[width=0.6\hsize]{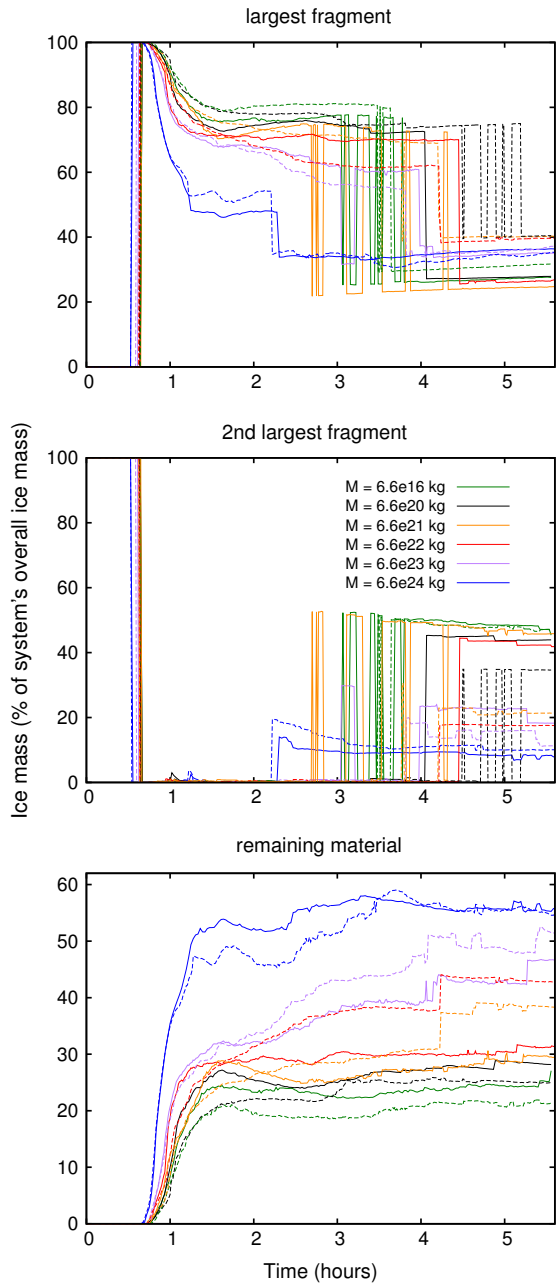}
	\caption{Evolution of the water ice masses of the largest fragment, the $2^\mathrm{nd}$ largest one, and the remaining material. Solid runs are indicated by solid lines and hydro runs by dashed lines.}
	\label{fig:fragment_masses_ice}
	\end{figure}
	\begin{figure}
	\centering
	\includegraphics[width=1.0\hsize]{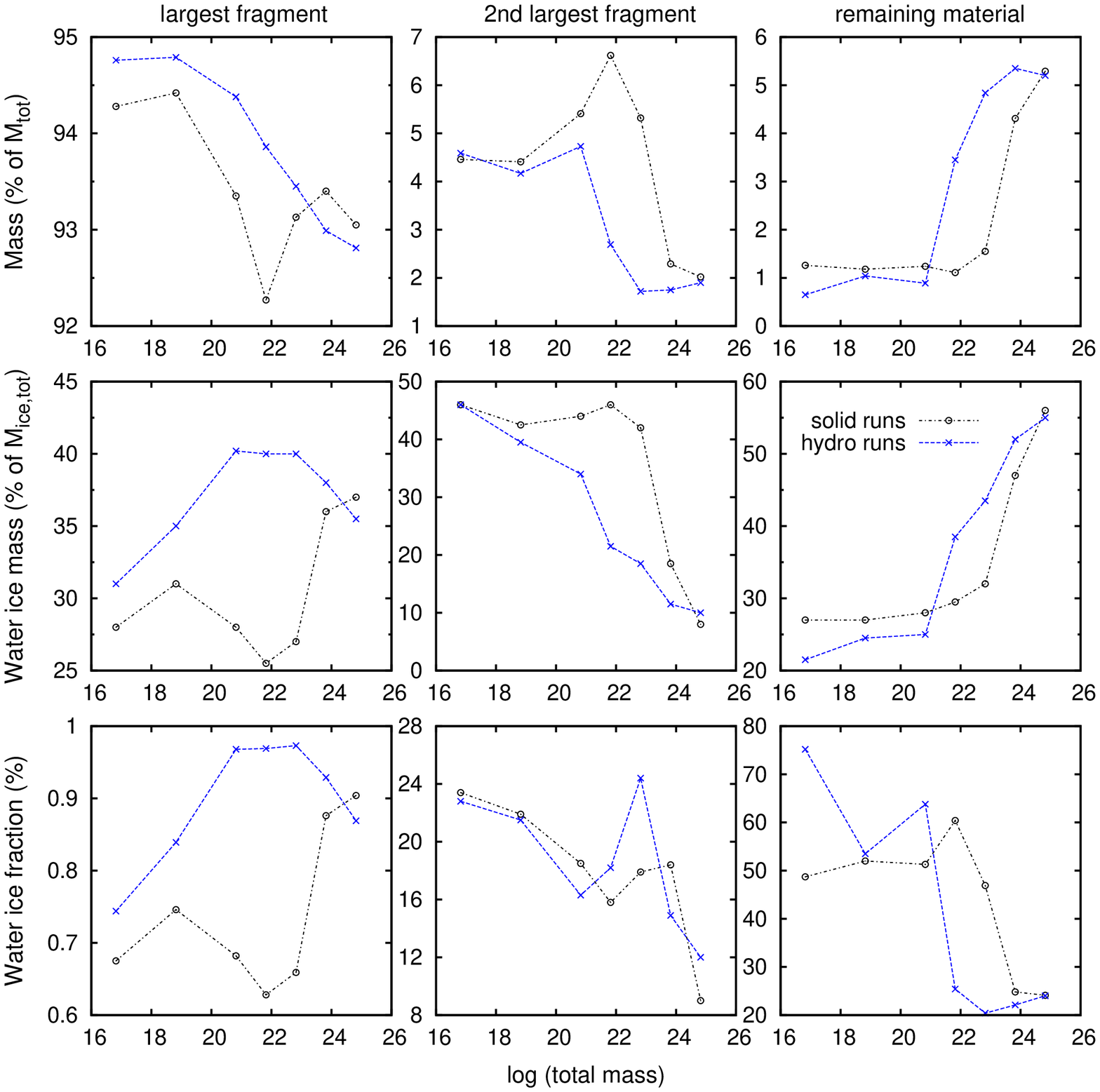}
	\caption{Final fragmentation results as a function of total scenario mass for all computed scenarios, showing overall fragment masses as fraction of the overall scenario mass (top row), their water ice masses as fraction of the scenarios' overall water ice mass (middle row) and water ice fractions of the individual fragments (bottom row). The data points in the upper 6 panels basically resemble the final states in the respective panels of Figs.~\ref{fig:fragment_masses_overall} and \ref{fig:fragment_masses_ice}.}
	\label{fig:final_fragments}
	\end{figure}
This general behavior gives rise to tracking these two largest fragments, as well as the combined rest of material.
The time evolution of the overall mass in these 3 categories is plotted in Fig.~\ref{fig:fragment_masses_overall}, while Fig.~\ref{fig:fragment_masses_ice} illustrates the development of water ice. The illustrations show all scenarios except for $\beta = 18$, whose results are close to the $\beta = 16$ and $\beta = 20$ scenarios, and are omitted for clarity in the plots.
Except for some fluctuations, which are caused by fragments briefly regaining contact (and loosing it again), the plots confirm the described behavior for all scenarios. After the initial approach the projectile and the target join into a single large body, followed by a chaotic interaction phase, before two main fragments (with respective water contents) emerge towards the end of simulated time. Even though there is still some minor accretion, the values in the rightmost parts of the curves can be considered sufficiently close to the theoretical final ones.
Figure~\ref{fig:final_fragments} provides a direct comparison of these final values of the solid and hydro runs.
The properties of real objects lie somewhere between these two models, depending on their size, thermal state and history. Even though collisions are complex processes they can thus somehow be thought of as limiting cases.
The first two panels in Fig.~\ref{fig:final_fragments} illustrate that the masses of the two largest fragments tend to decrease with total mass and the amount of remaining material tends to increase. However, for intermediate masses the solid runs strongly deviate from this trend.
The fate of water ice is even more diverse. From the amount carried initially by the projectile, the target (largest fragment) accretes between 25 and 40\%. The final water ice content of the $2^\mathrm{nd}$ largest fragment decreases from almost 50 to less than 10\% for the highest total mass, which directly correlates with this fragment's overall mass. The amount of lost water increases with total mass, from above 20 to about 55\%.

Approximative outbound velocities of the $2^\mathrm{nd}$ largest fragment directly after the actual collision are plotted in Fig.~\ref{fig:outbound_velocities}. They strongly increase with the colliding bodies' masses, from even below the escape velocity (i.e.\ they are actually gravitationally bound) to values very close to the projectile's inbound velocity before hitting the target (in units of $v_\mathrm{esc}$).

\section{Discussion and conclusions}
\label{sect:discussion_and_conclusions}
\begin{figure}
	\centering
	\includegraphics[width=1.0\hsize]{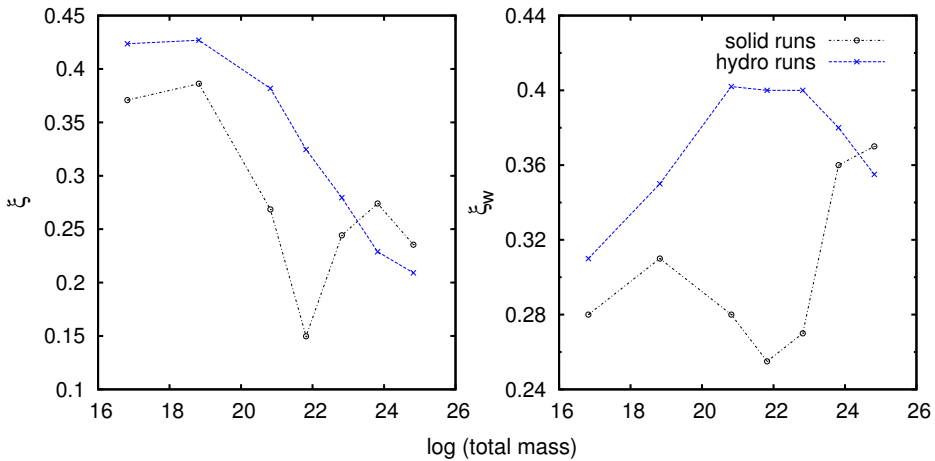}
	\caption{Overall accretion efficiency $\xi$ and water accretion efficiency $\xi_\mathrm{w}$ as a function of total scenario mass for all computed scenarios.}
	\label{fig:accretion_efficiencies}
	\end{figure}

The summarized results in Fig.~\ref{fig:final_fragments} indicate that there is indeed some degree of scale-invariance, at least for the very global outcome, over some, but not over the whole range of investigated masses. The principal similarity in simulation outcomes is also visible from Fig.~\ref{fig:16vs24}.
The largest fragment mass varies \emph{only} by about 2.5\% (Fig.~\ref{fig:final_fragments}), 
but expressing this in terms of the accretion efficiency, usually defined as
\begin{equation}
\xi = \frac{M_\mathrm{lf} - M_\mathrm{t}}{M_\mathrm{p}}\ ,
\label{eq:accretion_efficiency}
\end{equation}
with masses of the largest post-collision fragment $M_\mathrm{lf}$, the target and the projectile, yields values (Fig.~\ref{fig:accretion_efficiencies}) between $\xi = 0.15$ and $0.43$ for all runs, and still values between 0.2 and 0.43 if only hydro results are considered.
This means that the amount of projectile material accreted onto the target ranges from 15\% to above 40\%.
These results show significantly larger variations than those reported by \citet{Asphaug2010_Similar_sized_collisions_diversity_of_planets}, where they find $\xi$ to be scale-invariant within $\simeq 10$\% for differentiated terrestrial planets from Vesta to Super-Earth masses, even if their different mass range (starting at $\simeq 2.6\times 10^{20}$\,kg) is considered.
In addition we introduce a modified version of the classical accretion efficiency Eq.~(\ref{eq:accretion_efficiency}) to specifically capture transfer of water (ice)
\begin{equation}
\xi_\mathrm{w} = \frac{M_\mathrm{w,lf} - M_\mathrm{w,t}}{M_\mathrm{w,p}}\ ,
\label{eq:accretion_efficiency_water}
\end{equation}
with the water (ice) mass on the largest post-collision fragment $M_\mathrm{w,lf}$, the target and the projectile. In our scenarios the target is dry, therefore $\xi_\mathrm{w}$ gives the fraction of projectile water accreted by the target. As illustrated in Fig.~\ref{fig:accretion_efficiencies} this fraction varies relatively little -- between 25\% and 40\% -- considering the 10 orders of magnitude differences in mass. If both projectile and target initially contain water, $\xi_\mathrm{w}$ indicates water accretion (onto the target) for positive values, and water loss/erosion for negative ones, in units of the projectile's water mass $M_\mathrm{w,p}$. Note that $\xi_\mathrm{w}$ is not defined for a dry projectile.

As it is evident from Fig.~\ref{fig:final_fragments} the mass of the $2^\mathrm{nd}$ largest fragment and the remaining material seem to be relatively invariant of scale for masses up to $\simeq 10^{21}$\,kg, but vary strongly for higher masses.
If one is beyond that not only interested in the basic outcome but in more detailed behavior -- like transfer and loss of a water ice mantle -- the assumption of scale-invariance is certainly not met in general, at least not over a broad range of masses.
As mentioned above the target accretes between 25 and 40\% of all water ice from the projectile, the rest either remains on the $2^\mathrm{nd}$ largest fragment (or is reaccreted by it) or is lost. 
Interestingly the (initially dry) target ends up with a water mass fraction somewhere between 0.6 and 1\%, which is (roughly) comparable to the $\simeq 0.1$\% \citep{Dishoeck2014_Water_from_clouds} of current Earth.
The trend that water ice contents of the $2^\mathrm{nd}$ largest fragment strongly decrease towards high masses (middle-middle panel in Fig.~\ref{fig:final_fragments}) is not necessarily due to the genuine properties of water, but connected to the lower overall masses of this fragment towards the highest masses. However, the water ice fraction of the $2^\mathrm{nd}$ largest fragment is indeed -- mostly -- a decreasing function of mass (lower-middle panel in Fig.~\ref{fig:final_fragments}).
It was suggested that collisions similar to our scenarios can strip most of the projectile's mantle, particularly volatile material \citep[e.g.][]{Asphaug2010_Similar_sized_collisions_diversity_of_planets}. This is only partly the case for our results, where the $2^\mathrm{nd}$ largest fragment's water ice fraction is even close to the projectile's pre-collision value of 25\% for low-mass scenarios, remains around 15 to 20\% for most outcomes, and does not drop below $\simeq$10\% for the highest masses.

For the material lost as debris (the rightmost panels in Fig.~\ref{fig:final_fragments}) interestingly this figure is not only strongly increasing towards the highest masses, but also its composition changes. The projectile has 1/11 of the total mass and a water ice fraction of 0.25, hence the overall water ice fraction w.r.t.\ $M_\mathrm{tot}$ is $1/44 \simeq 2.3$\%. For the low-mass scenarios 50\% or more of the lost material is water ice, while for the most-massive ones the material ratio is up to 4:1 for basalt, likely a result of high impact energies and therefore more violent encounters, which particularly affect (and subsequently disperse) also the projectile's basalt interior (cf.\ Fig.~\ref{fig:16vs24}).

	\begin{figure}
	\centering
	\includegraphics[width=0.43\hsize]{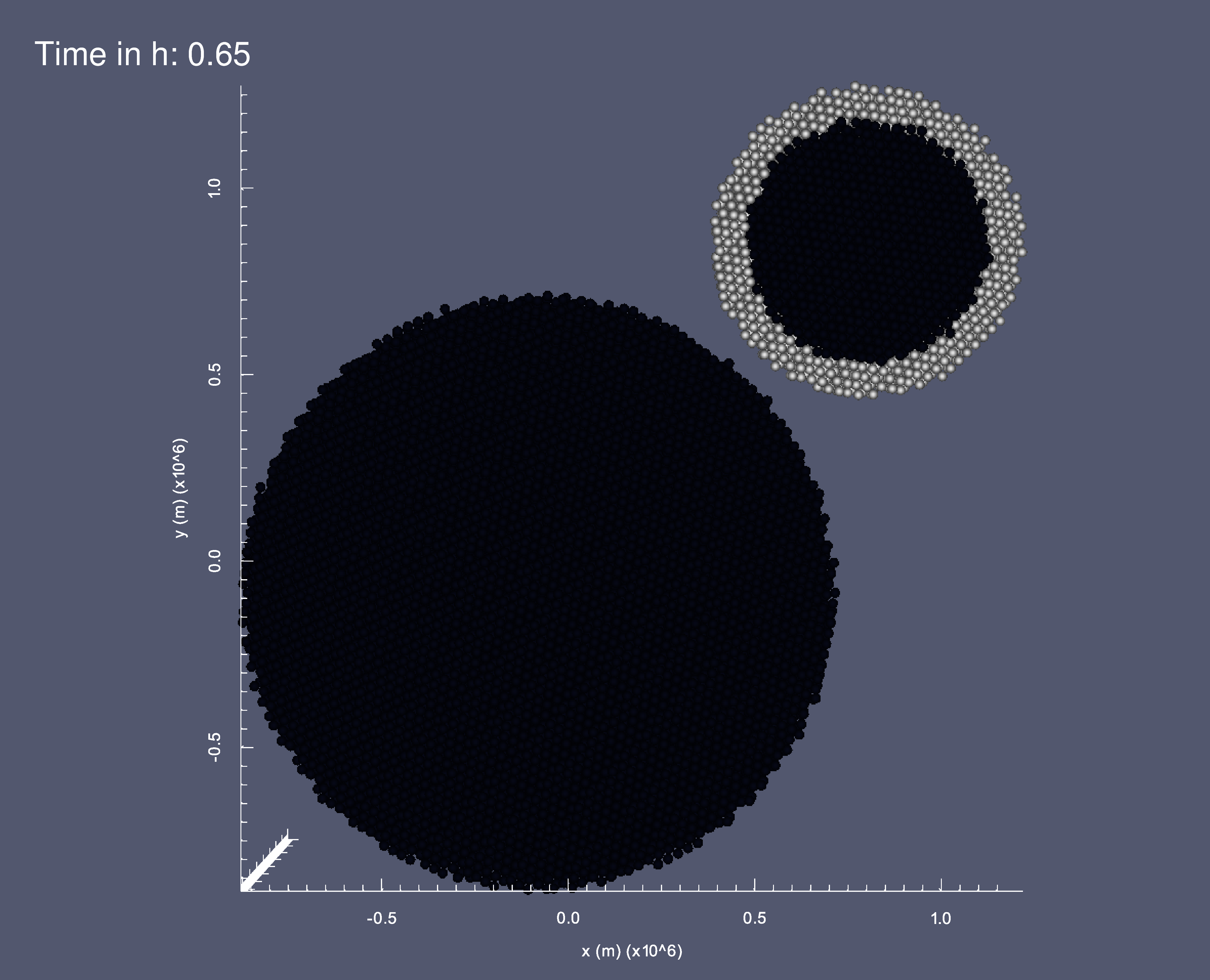}
	\includegraphics[width=0.43\hsize]{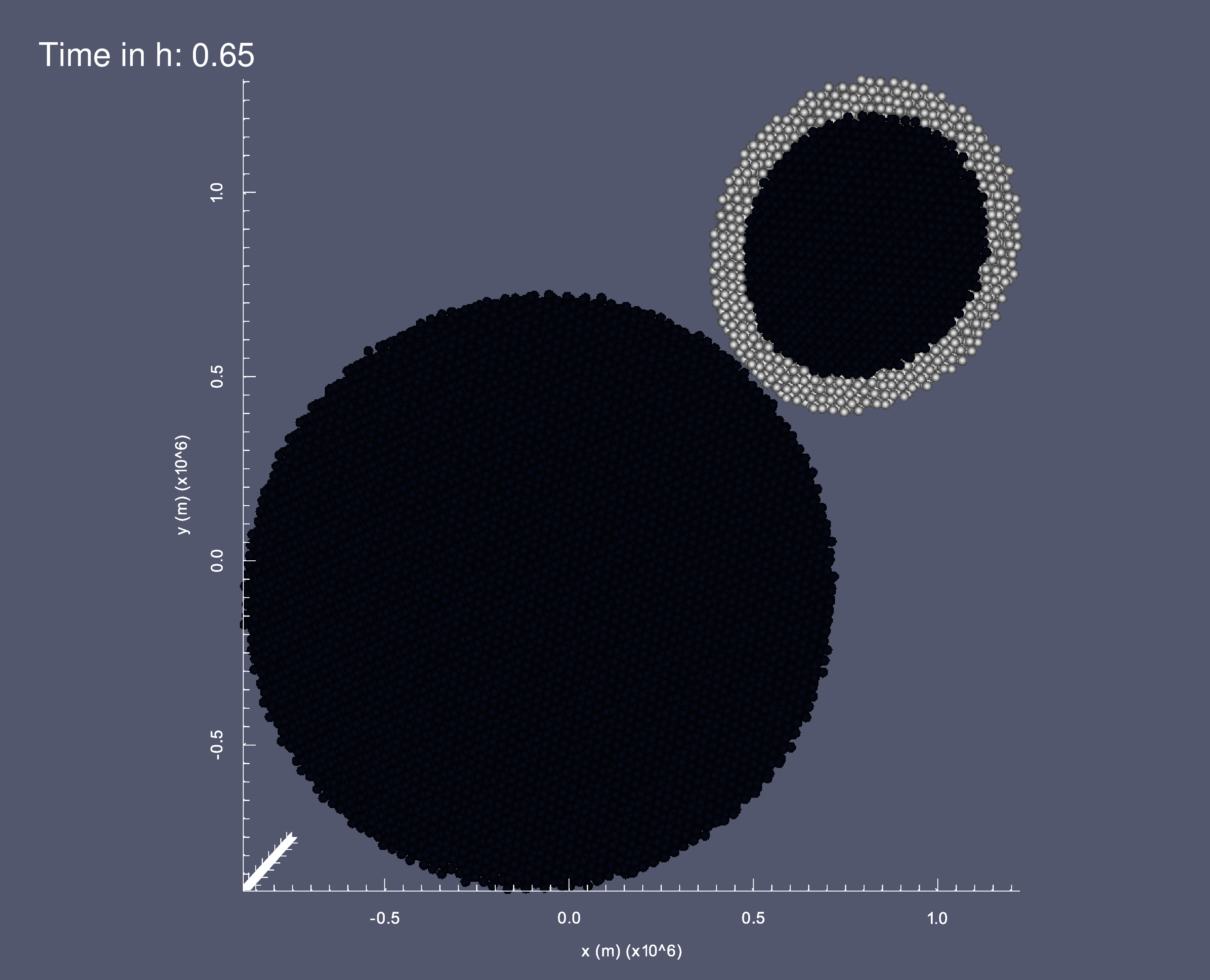} \\
	\vspace{0.1mm}
	\includegraphics[width=0.43\hsize]{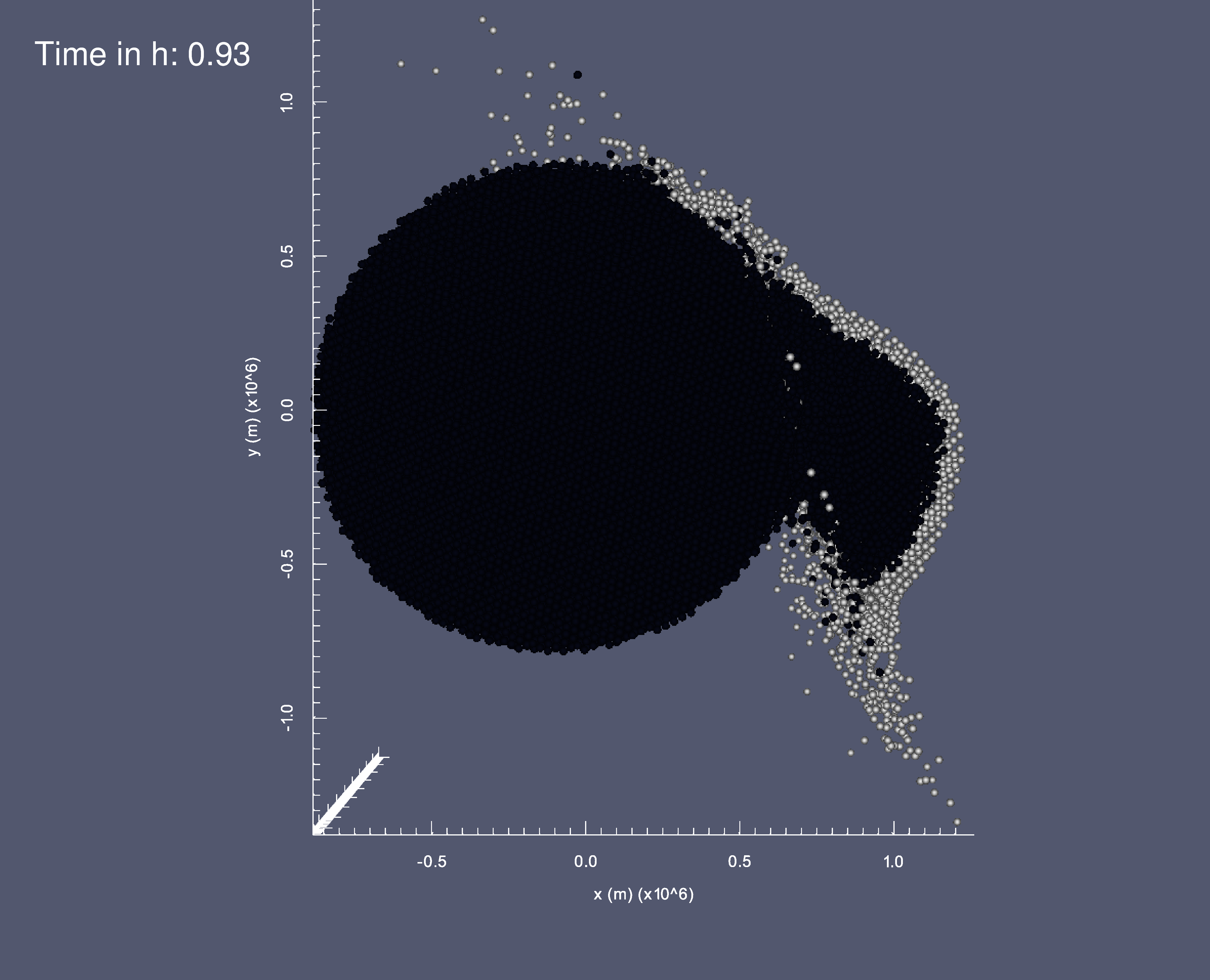}
	\includegraphics[width=0.43\hsize]{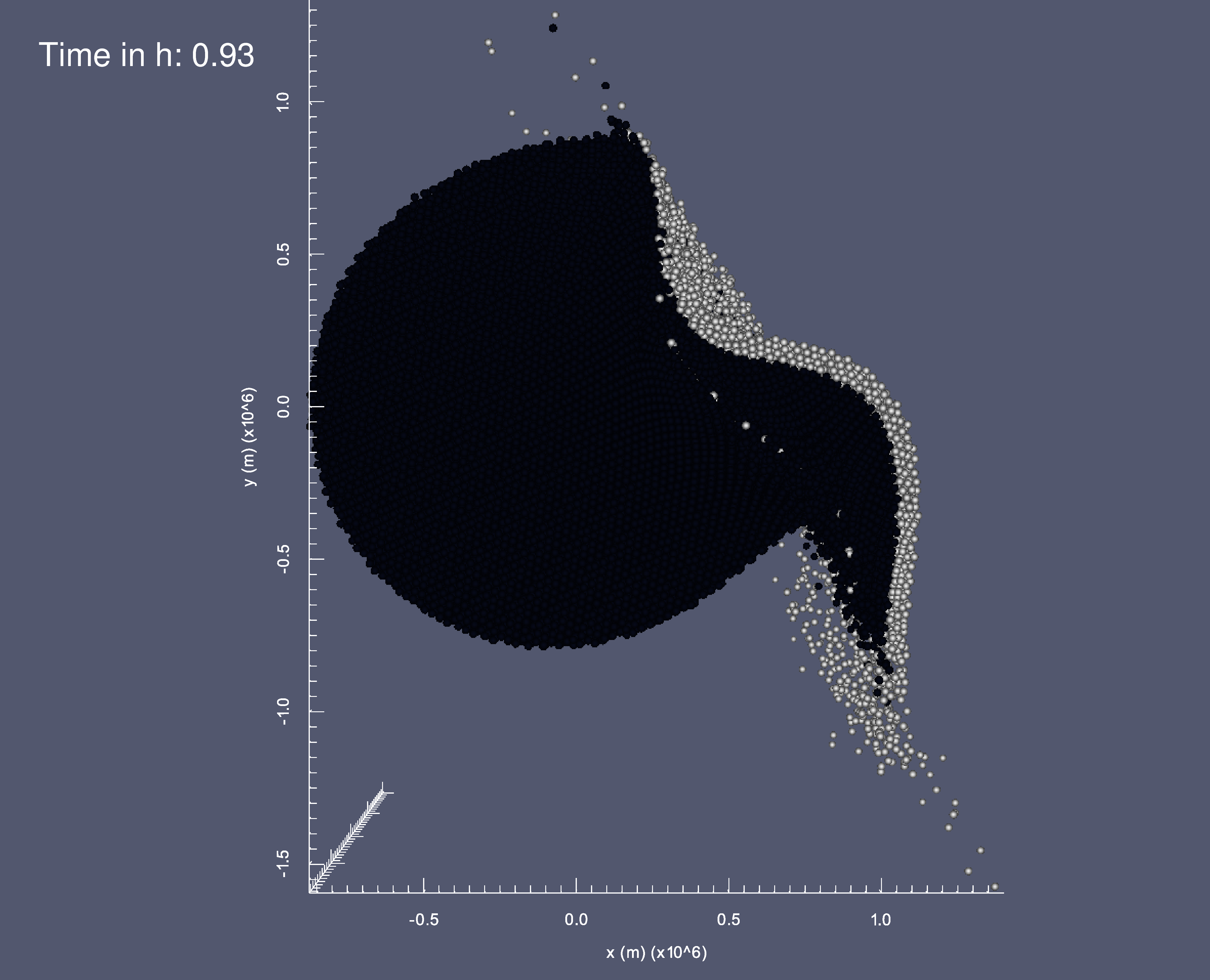} \\
	\vspace{0.1mm}
	\includegraphics[width=0.43\hsize]{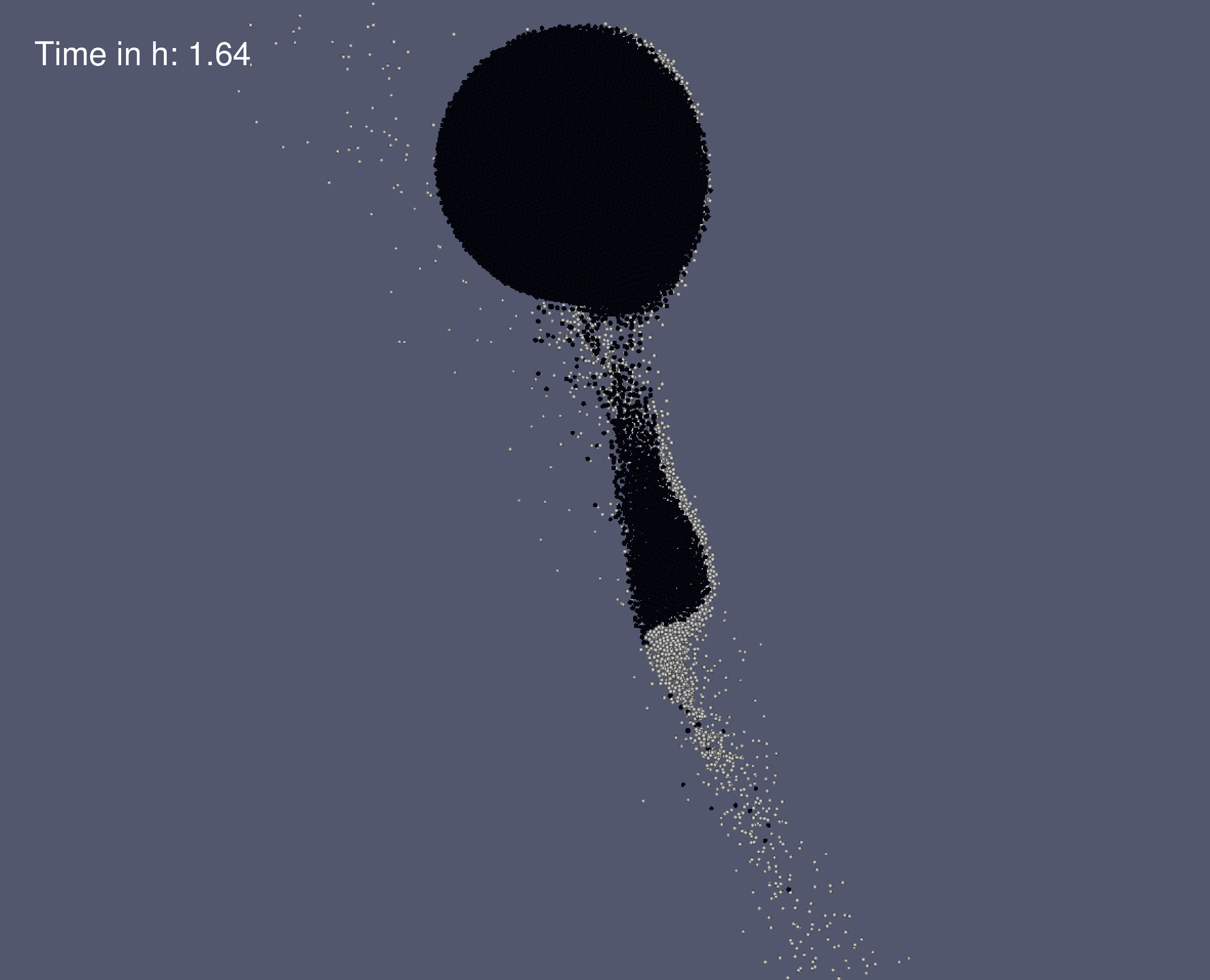}
	\includegraphics[width=0.43\hsize]{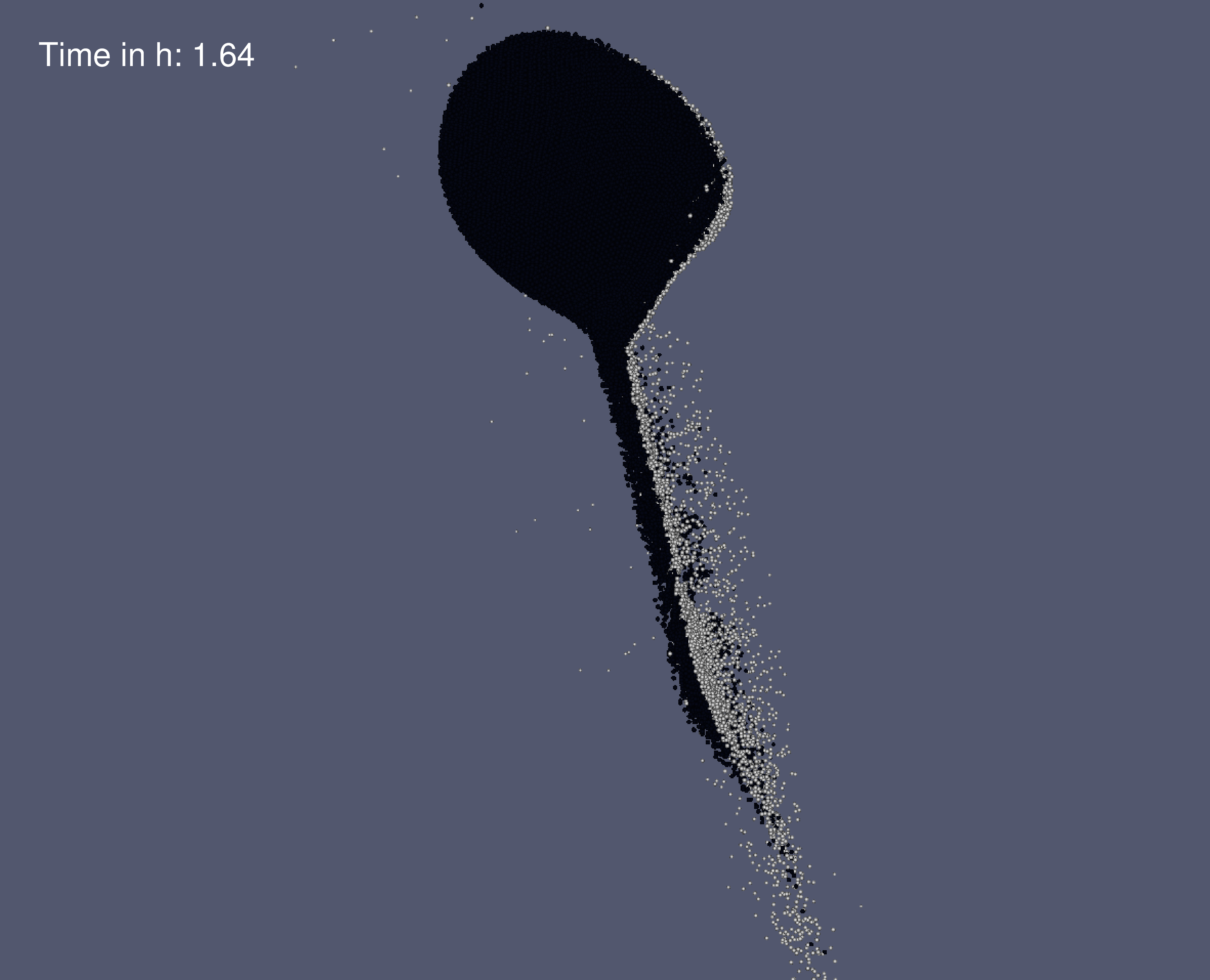} \\
	\vspace{0.1mm}
	\includegraphics[width=0.43\hsize]{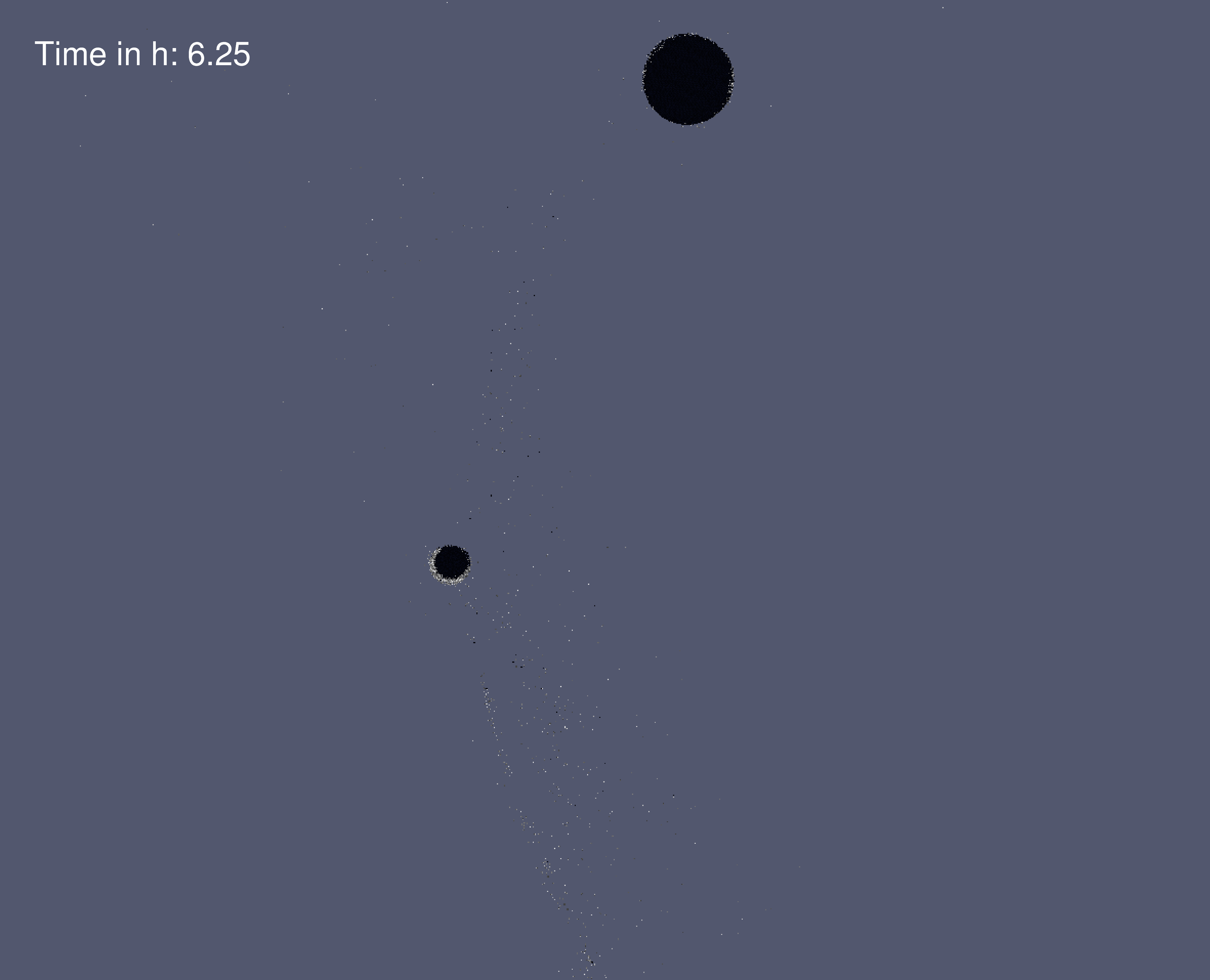}
	\includegraphics[width=0.43\hsize]{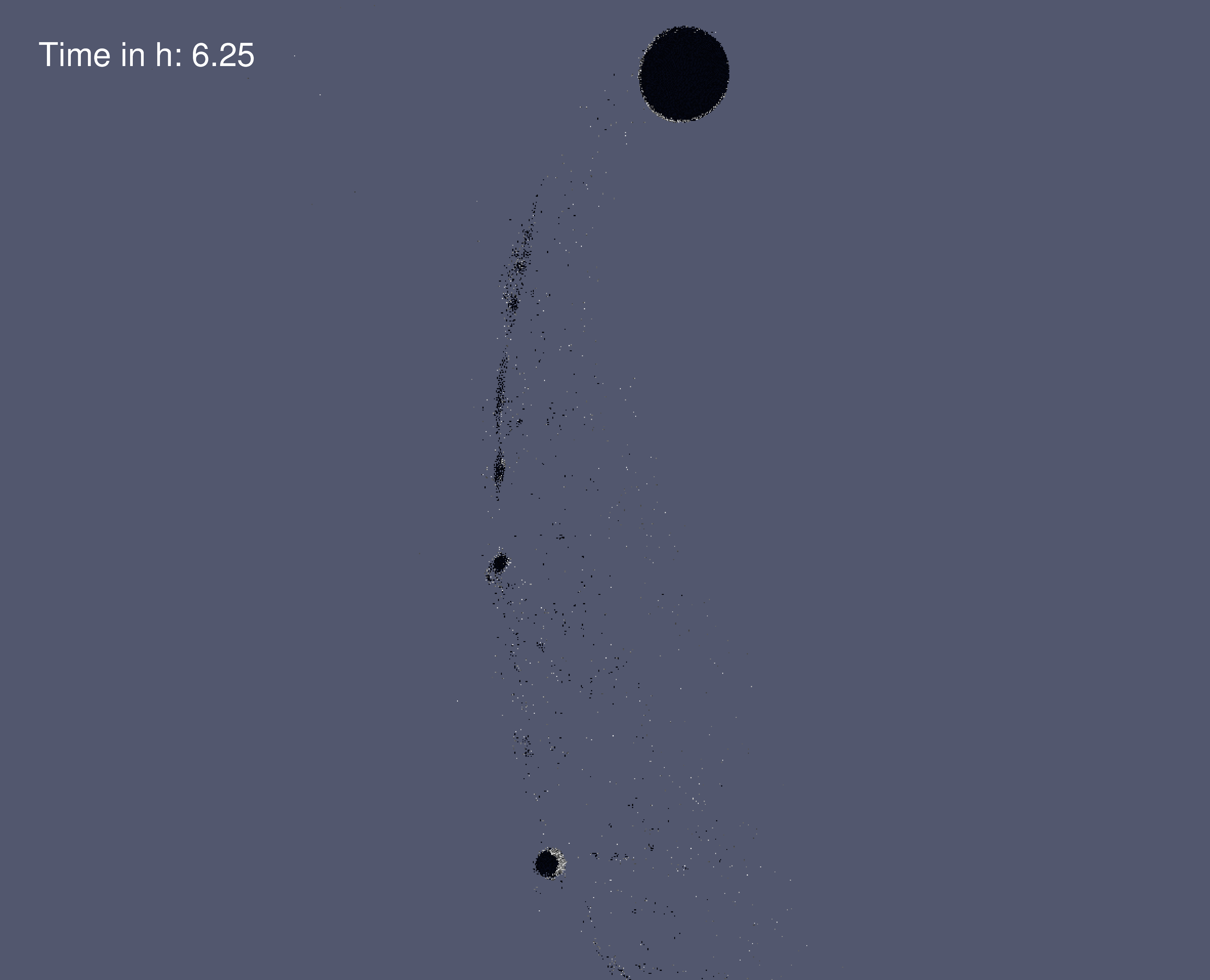}
	\caption{Comparison of simulation snapshots (bodies cut into halves) of the solid (left) and hydro (right) runs of the $\beta = 21$ scenario. Material types are color-coded in black (basalt) and white (water ice). The respective solid/hydro snapshots show exactly equal times.}
	\label{fig:solid_hydro}
	\end{figure}
A remarkable result are the differences between solid and hydro runs, which are interestingly most pronounced in the intermediate mass scenarios (cf. Fig.~\ref{fig:final_fragments}). We illustrate this behavior by comparing simulation snapshots of a solid to a hydro run in Fig.~\ref{fig:solid_hydro}.
The material strength (tensile and shear stresses) included in the solid run strongly decelerate and also deflect (to the right) the projectile upon impact, while these effects are much less pronounced for the hydro run, where the projectile can easily penetrate the target and the visual impression is much more \emph{liquid}.
After the actual collision the projectile material remains more compact and also significantly slower in the solid run. This is expressed in later times by a single big $2^\mathrm{nd}$ largest fragment in the solid case, but a smaller and faster one in the hydro run, along with a lot more debris (cf.\ Fig.~\ref{fig:final_fragments}).
Since the initial projectile remains much more intact in the solid run it can drag more of its original water ice content with it, which explains the greater amount of water ice on the $2^\mathrm{nd}$ largest fragment (but not a greater water ice fraction) and the lower amount of transferred water ice to the target (and also a significantly lower water ice fraction).
Another subtle difference is the build-up of much stronger tidal deformation prior to the impact in the hydro run due to absent material strength (see Fig.~\ref{fig:solid_hydro}).

	\begin{figure}
	\centering
	\includegraphics[width=0.49\hsize]{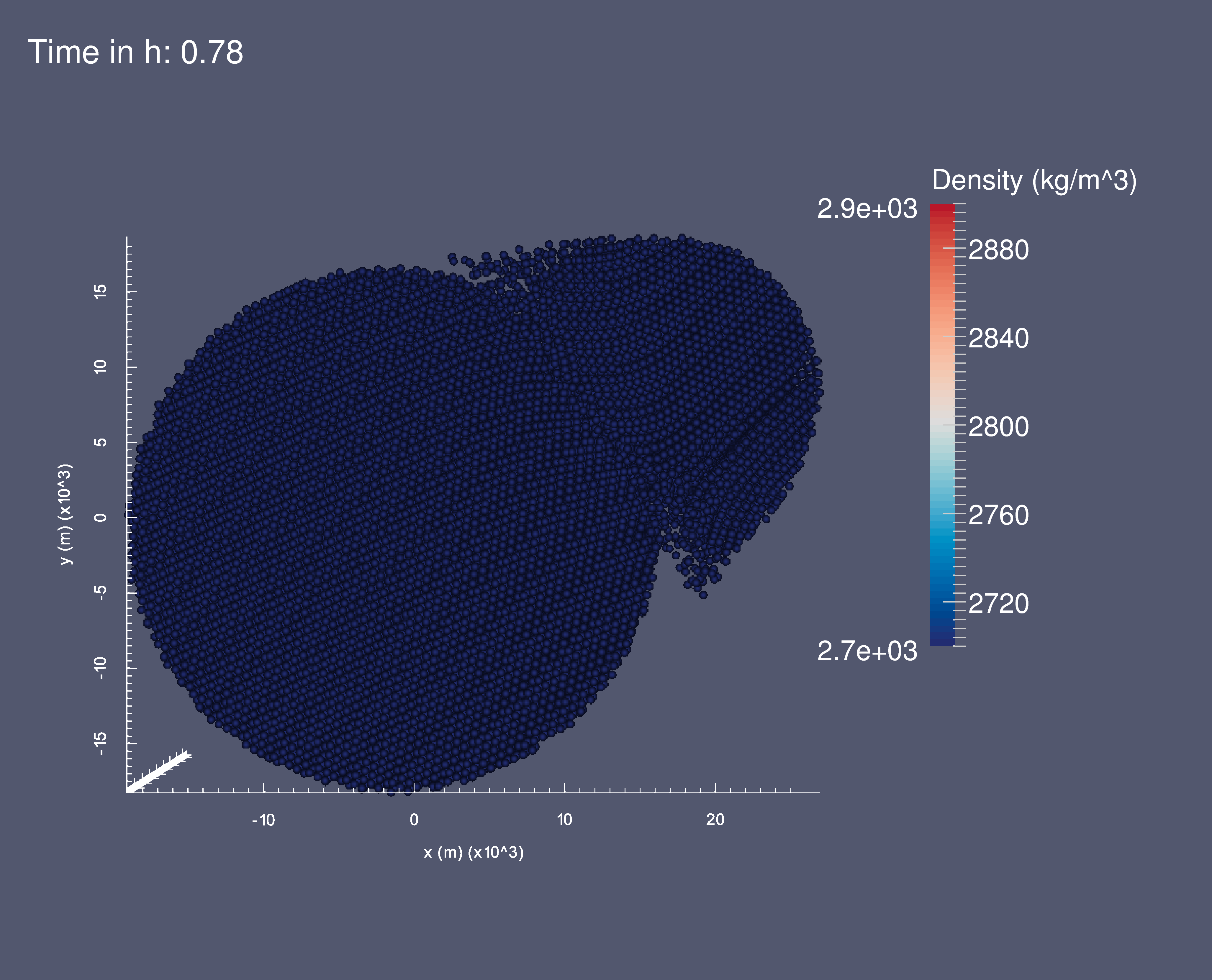}
	\includegraphics[width=0.49\hsize]{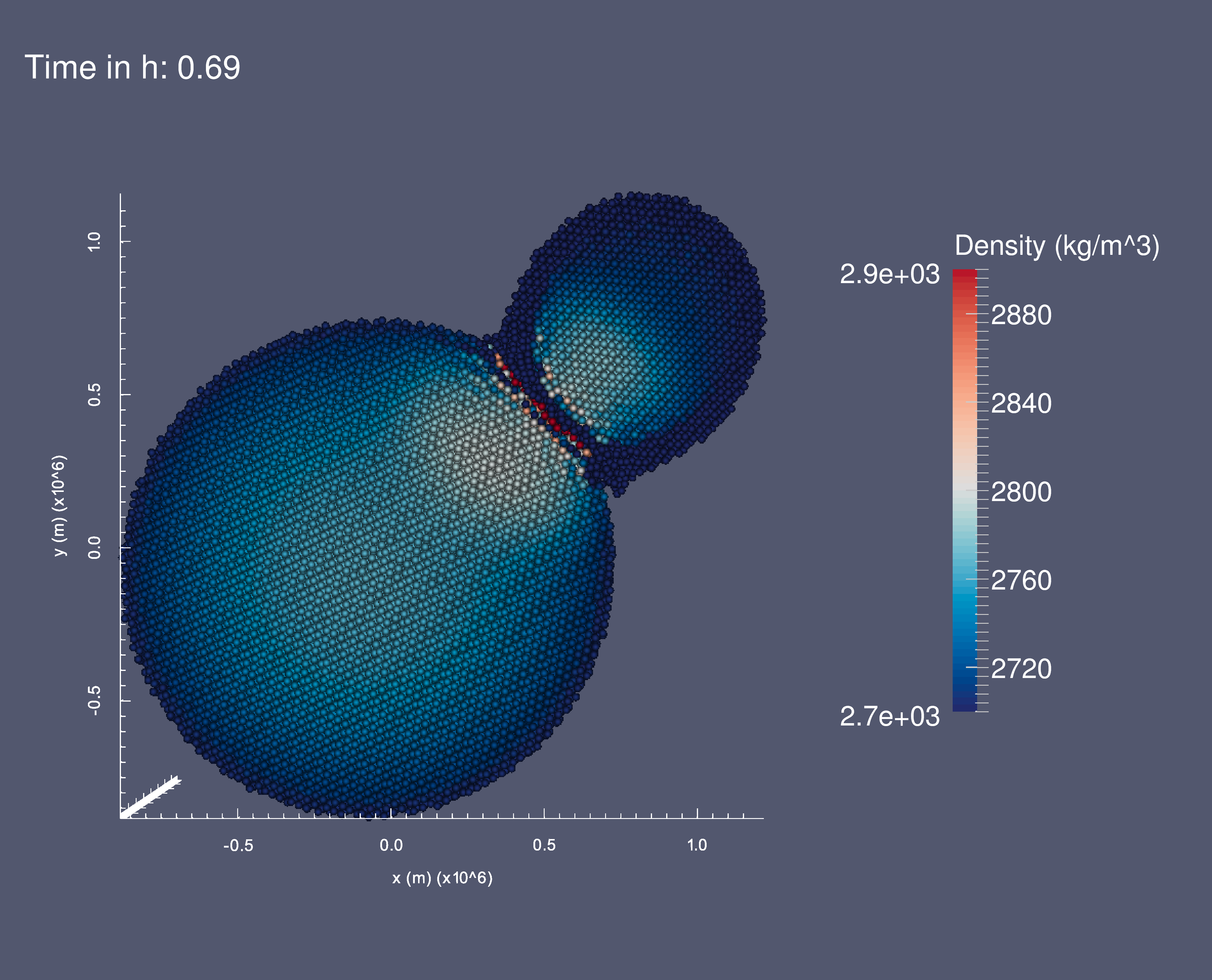} \\
	\vspace{0.3mm}
	\includegraphics[width=0.49\hsize]{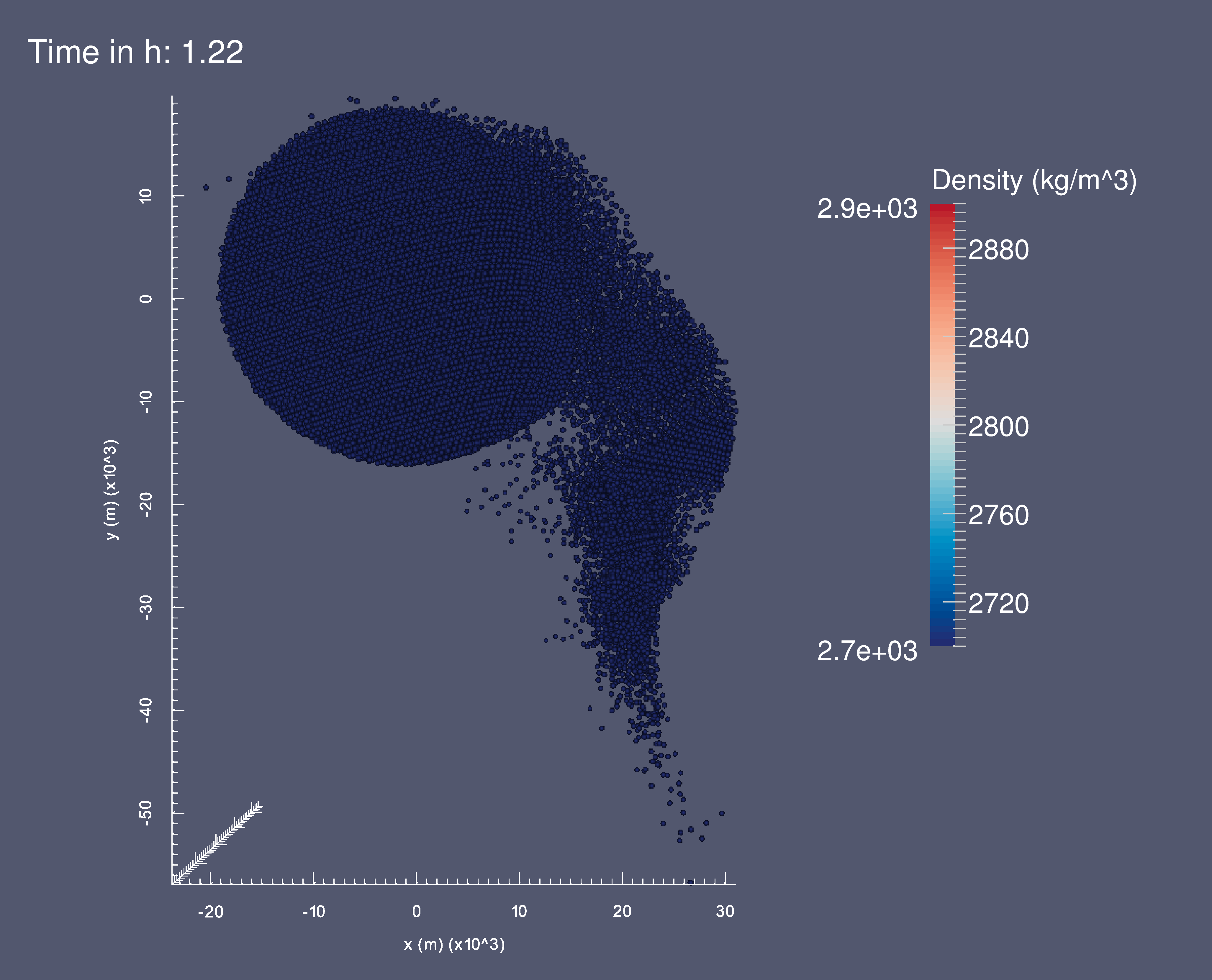}
	\includegraphics[width=0.49\hsize]{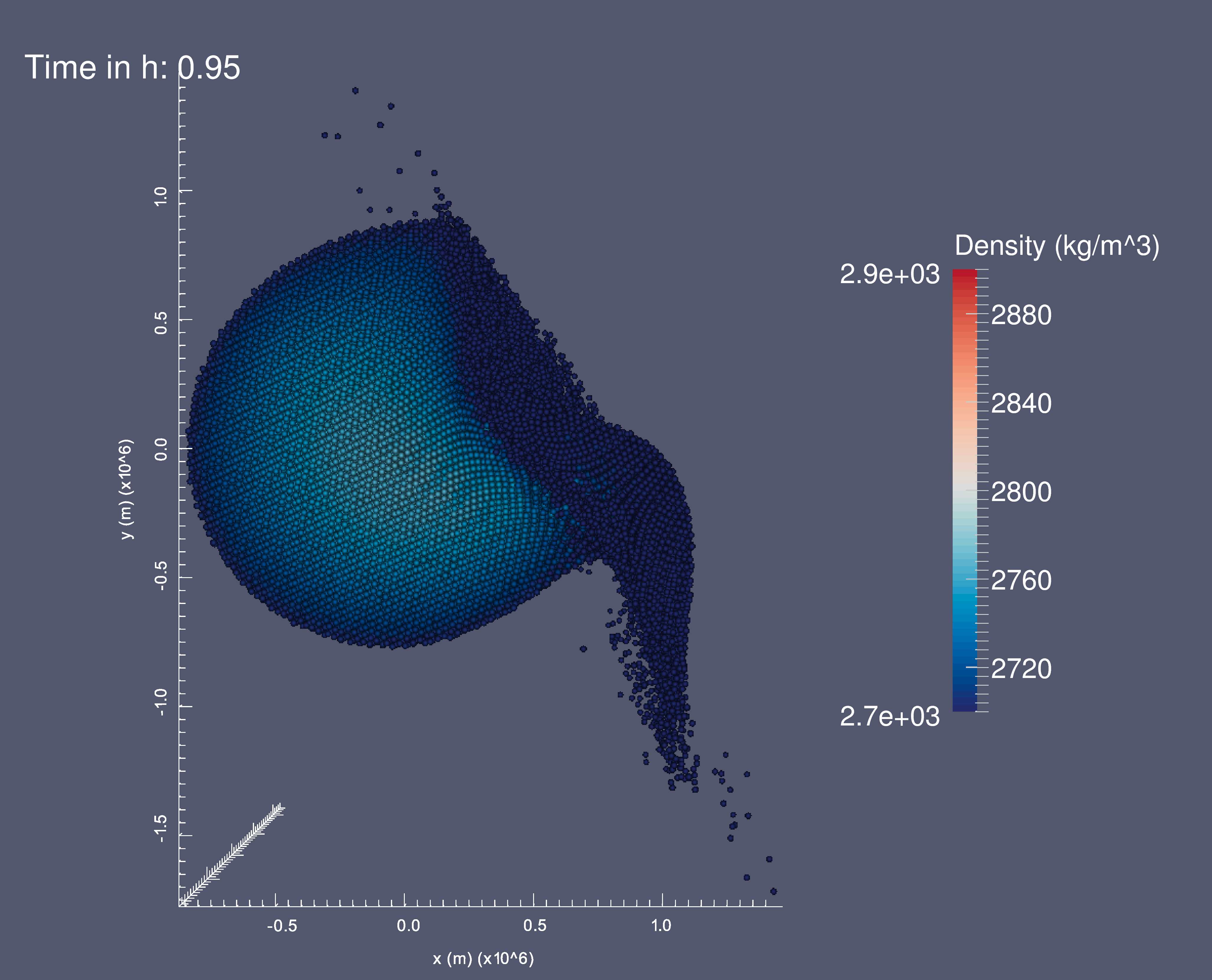}
	\caption{Comparison of simulation snapshots (bodies cut into halves) of the hydro runs of the $\beta = 16$ scenario (left) and the $\beta = 21$ scenario (right). The density distribution is color-coded, where basalt has an uncompressed density of $2700\,\mathrm{kg/m^3}$.}
	\label{fig:hydro_comparison}
	\end{figure}
\begin{figure}
	\centering
	\includegraphics[width=0.75\hsize]{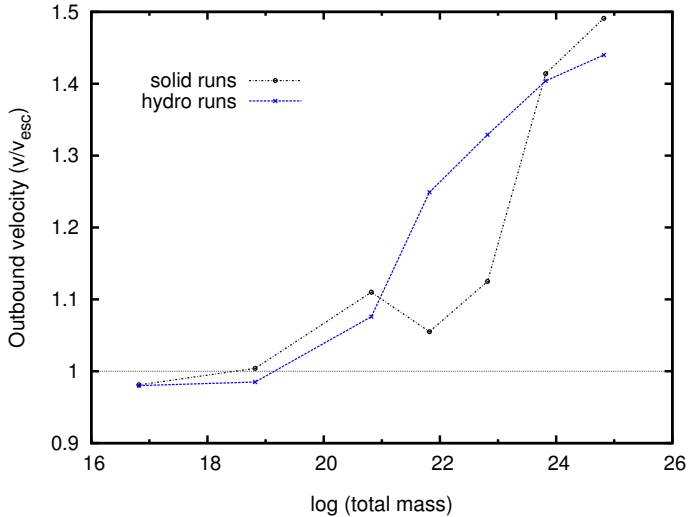}
	\caption{(Theoretical) outbound velocities of the $2^\mathrm{nd}$ largest fragment w.r.t. the most-massive one at $R_\mathrm{P}+R_\mathrm{T}$ for it to end up at its final position in the last simulation output frame.}
	\label{fig:outbound_velocities}
	\end{figure}

Why is the difference between solid and hydro runs most emphasized for the intermediate mass scenarios, and not increasing with decreasing total mass -- where material strength generally becomes more important because impact energies are lower?
For our low-mass scenarios impact energies are too small to significantly compress material, therefore the projectile is falling/flowing apart when it (slowly, tens of m/s) impacts the (for it) incompressible target, instead of compressing and penetrating target material as it happens for higher impact energies (and thus masses for our scenario setup). This behavior is illustrated in Fig.~\ref{fig:hydro_comparison}, by comparing the hydro runs of the $\beta = 16$ and $\beta = 21$ scenarios. While for low impact energies almost no compression can happen and the projectile is diverted (left panels), the situation is clearly different for higher energies and thus pressures. A comparison of the $\beta = 16$ panels in Fig.~\ref{fig:hydro_comparison} and those in Fig.~\ref{fig:16vs24} illustrates the similarity between low-mass solid and hydro runs, caused by this behavior.
Therefore it seems that for our lowest mass scenarios impact energies are too small to cause significant compression and thus displacement of material, which also results in similar results for solid and hydro runs. This also leads to more dissipation of kinetic energy in the target and as a consequence lower outbound fragment velocities (see below) and lower amounts of debris (cf.\ Fig.~\ref{fig:final_fragments}). When masses increase -- specifically our intermediate mass scenarios $\beta = 21,22$ -- it becomes decisive whether bodies have strength or behave like strengthless fluids (Fig.~\ref{fig:solid_hydro}). For solid runs tensile and shear stresses result in similar outcomes to the low-mass scenarios, while results for strengthless bodies are closer to the high-mass scenarios.
Finally, when approaching planetary masses for scenarios $\beta=23$ and 24 here, high impact energies and stresses result in strong compression of target (and also projectile) material, and material strength is not important anymore -- solid and hydro runs have similar outcomes. The projectile experiences little resistance by the target material, fragments are strongly dispersed and their velocities are high (Fig.~\ref{fig:final_fragments} and \ref{fig:outbound_velocities}).
Bodies on these and larger scales posses highly compressed interiors, resulting in increased bulk densities. A consequence is an inherent deviation from scale invariance caused by impact velocities that are higher than would be for uncompressed bodies ($v_\mathrm{esc}$ is higher), but the projectile still ploughs through the relatively uncompressed outer layers of the target in grazing collisions like ours.
An additional potential influence is due to shocks. The speed of sound in the simulated materials is around 3\,km/s, for our scenarios this transition is between $\beta=22$ (slightly subsonic) and $\beta=23$ (clearly supersonic).

Scaling also significantly affects the dynamics. For most scenarios except the most-massive ones the initial approach of projectile and target is almost scale-invariant, i.e. it takes them the same time until collision and they have the same $v/v_\mathrm{esc}$ for equal times (all scenarios start at a distance of $3\times (R_\mathrm{P}+R_\mathrm{T})$\,). Due to their strong hydrostatic compression the most-massive scenarios deviate from this behavior -- their radii are significantly smaller than would be for homogeneous density, and therefore they start closer (but still at $3\times (R_\mathrm{P}+R_\mathrm{T})$\,) and collide earlier (visible in Fig.~\ref{fig:16vs24}).
Fragment dynamics after the actual collision are complicated, diverse and hard to compare due to differing fragmentation behavior of the scenarios. We will not elaborate on all the details, but to have some comparison consider the relative orbit of the $2^\mathrm{nd}$ largest fragment and the largest one after the collision. If one takes their relative positions and velocities in the last simulation output frame and computes their 2-body orbit backwards, say to a distance $R_\mathrm{P}+R_\mathrm{T}$, the result is the outbound velocity the $2^\mathrm{nd}$ largest fragment would have had to have directly after the encounter with the target (neglecting other material).
Fig.~\ref{fig:outbound_velocities} illustrates that this velocity (in units of $v_\mathrm{esc}$) increases with total mass, and furthermore that for the lowest mass scenarios the outcome is actually entirely different than for the other scenarios, with gravitationally bound $2^\mathrm{nd}$ largest fragments.
While the unbound cases represent collisions between partial accretion and hit-and-run \citep{Stewart2012_Collisions_between_gravity-dominated_bodies_II}, the scenarios with a gravitationally bound $2^\mathrm{nd}$ largest fragment fall into the graze-and-merge regime (the two largest fragments are assumed to eventually merge).
Even though these results are biased by the differing masses of the $2^\mathrm{nd}$ largest fragment, it illustrates the transition from low-mass to high-mass collisions.

The scenarios considered here represent only one point in a large, multi-dimensional parameter space, consisting at least of the projectile-to-target mass ratio, the impact angle and the impact velocity.
Therefore one has to be careful when mapping them to sufficiently different collision geometries.
Finally it depends on the level of detail one is interested in and particularly on the broadness of the considered mass range whether scale-invariance can be approximately assumed or not. However, care has to be taken, and simply relying on scale-invariance without testing it could render results unreliable.
An application of giant collision simulations is to model fragmentation behavior in N-body computations of late-stage planet formation, and to possibly track retention and loss of volatiles (water) in future studies. Current N-body simulations of that kind usually start with planetesimal/embryo masses between Moon- ($\simeq 7\times 10^{22}$\,kg) and Mars-sized ($\simeq 6\times 10^{23}$\,kg) bodies.
Intermediate mass hydro runs behave relatively similar to high mass runs, because of the absence of material strength, therefore approximative scale-invariance may be assumed over a range of masses typically used in current N-body simulations, for sufficiently strengthless bodies -- which might be a reasonable assumption for large bodies in an active planet formation environment.

We did not expect (simple) scale-invariance over this broad range of masses when we started working on this study, but were still surprised by the results' broad diversity. The intention was to investigate exemplarily how collision outcomes change with scale (and material model). 
To account for the scale effects encountered throughout this study the usual approach is to identify relations that predict some aspect of a collision -- like fragment masses, compositions or velocities -- for varying scales, referred to as scaling laws.
Extensive literature on this topic exists \citep{HousenHolsapple1990_Fragmentation_of_asteroids_and_satellites, Holsapple1993_Scaling_of_impact_processes, Marcus2010_Icy_Super_Earths_max_water_content, Leinhardt2012_Collisions_between_gravity-dominated_bodies_I, Stewart2012_Collisions_between_gravity-dominated_bodies_II, Movshovitz2016_Impact_disruption_new_sims_and_scaling}, with this (incomplete) list containing preferentially work relevant for similar-sized collisions.
Ultimately this means that not the mass ratio, $v/v_\mathrm{esc}$ and impact angle need to be identical to produce similar outcomes, but rather some more complicated quantity, where this supposed quantity obviously differs between our scenarios.

\acknowledgements{We want to thank Rudolf Dvorak and Harry Varvoglis for organizing this very interesting workshop, and also Thomas Maindl for valuable comments on the manuscript. C.\,B. acknowledges support by the FWF Austrian Science Fund project S11603-N16.}

\bibliographystyle{aasjournal}
\bibliography{./additional_references,./references_EOS,./references_SPH,./references_Planet_formation,./references_Water}
\clearpage

\end{document}